\documentclass[12pt]{article}
\usepackage[latin9]{inputenc}
\usepackage[a4paper]{geometry}
\usepackage{float}
\usepackage{mathtools}
\usepackage{amsmath}
\usepackage{amsthm}
\usepackage{amssymb}
\usepackage{esint}
\usepackage{pifont}
\usepackage[numbers,sort&compress]{natbib}
\usepackage{color}
\usepackage[unicode=true,
 bookmarks=true,bookmarksnumbered=false,bookmarksopen=false,
 breaklinks=false,pdfborder={0 0 0},pdfborderstyle={},backref=false,colorlinks=false]
 {hyperref}

\def\bal#1\eal{\begin{align}#1\end{align}}
\def\alp[#1]{\begin{align}#1\end{align}}

\def\secnum[#1]{\texorpdfstring{$#1$}{TEXT}}

\def\secnuml#1\secnumr{\texorpdfstring{$#1$}{TEXT}}

\def\eqa{\begin{eqnarray}}
\def\eqae{\end{eqnarray}}
\def\eq{\begin{equation}}
\def\eqe{\end{equation}}
\def\be{\begin{equation}}
\def\ee{\end{equation}}
\def\bea{\begin{eqnarray}}
\def\eea{\end{eqnarray}}
\def\ba{\begin{array}}
\def\ea{\end{array}}
\def\bd{\begin{displaymath}}
\def\ed{\end{displaymath}}

\def\>{\rangle}
\def\<{\langle}

\makeatletter
\numberwithin{equation}{section}
\numberwithin{figure}{section}
\theoremstyle{plain}

\theoremstyle{definition}

\theoremstyle{plain}

\theoremstyle{plain}

\theoremstyle{plain}

\usepackage{braket}
\usepackage{feynmf}
\usepackage{subfigure}

\makeatother

\providecommand{\corollaryname}{Corollary}
\providecommand{\definitionname}{Definition}
\providecommand{\lemmaname}{Lemma}
\providecommand{\propositionname}{Proposition}
\providecommand{\theoremname}{Theorem}

\begin{document}


\begin{titlepage}

\thispagestyle{empty}

\begin{flushright}
\end{flushright}

\vspace{.4cm}
\begin{center}
\noindent{\large \bf  Page Curve from Defect Extremal Surface\\ and Island in Higher Dimensions}\\
\vspace{2cm}
Jinwei Chu$^{13}$, Feiyu Deng$^1$, Yang Zhou$^{12}$
\vspace{1cm}

{\it
$^1$ Department of Physics,
Fudan University, Shanghai 200433, China\\
$^2$ Peng Huanwu Center for Fundamental Theory, Hefei, Anhui 230026, China\\
$^3$ Department of Physics, University of Chicago, Chicago, Illinois 60637, USA
}

\end{center}

\vspace{.5cm}
\begin{abstract}
Defect extremal surface is defined by minimizing the Ryu-Takayanagi surface corrected by the defect theory, which is useful when the RT surface crosses or terminates on the defect. Based on the decomposition procedure of a AdS bulk with a defect brane, proposed in arXiv:2012.07612, we derive Page curve in a time dependent set up of AdS$_3$/BCFT$_2$, and find that the result from island formula agrees with defect extremal surface formula precisely. We then extend the study to higher dimensions and find that the entropy computed from bulk defect extremal surface is generally less than that from island formula in boundary low energy effective theory, which implies that the UV completion of island formula gives a smaller entropy in higher dimensions.
\end{abstract}

\end{titlepage}

\setcounter{tocdepth}{3}
{\hypersetup{linkcolor=black}\tableofcontents}

\newpage

\section{Introduction}
Black hole information paradox is a problem over 40 years. Recent progress hints towards a new understanding of the late time black hole interior as part of the Hawking radiation, which is called island. In particular the island formula for the radiation entropy gives Page curve~\cite{Page:1993wv,Page:2013dx,Hawking:1976ra} and therefore maintains unitarity. The key step to reproduce Page curve in recent breakthrough works~\cite{Penington:2019npb,Almheiri:2019psf} is to employ the quantum extremal surface (QES) formula for the fine grained entropy, which was inspired from the quantum corrected Ryu-Takayanagi formula in computing holographic entanglement entropy~\cite{RT:RT-formula,HRT:HRT-formula,Faulkner:2013ana,Engelhardt:2014gca}. For recent related works, see \cite{Almheiri:2019yqk,Almheiri:2019qdq,Almheiri:2019hni,Penington:2019kki,Rozali:2019day,Chen:2019iro,Verlinde:2020upt,
Chen:2020wiq,Gautason:2020tmk,Anegawa:2020ezn,Giddings:2020yes,
Hashimoto:2020cas,Sully:2020pza,Hartman:2020swn,Hollowood:2020cou,
Alishahiha:2020qza,Geng:2020qvw,Zhao:2019nxk,Chen:2019uhq,Almheiri:2019psy,Li:2020ceg,
Chandrasekaran:2020qtn,Bak:2020enw,Bousso:2020kmy,Dong:2020uxp,Balasubramanian:2020jhl,
Chen:2020jvn,Chen:2020tes,Emparan:2020znc,Hartman:2020khs,Murdia:2020iac,VanRaamsdonk:2020tlr,Liu:2020jsv,Langhoff:2020jqa,Balasubramanian:2020xqf,Balasubramanian:2020coy,
Grado-White:2020wlb,Sybesma:2020fxg,Mirbabayi:2020fyk,Chen:2020uac,Chen:2020hmv,Ling:2020laa,Bhattacharya:2020uun,Marolf:2020rpm,Harlow:2020bee,Nomura:2020ska,Hernandez:2020nem,Chen:2020ojn,Kirklin:2020zic,Matsuo:2020ypv,Goto:2020wnk,Hsin:2020mfa,Akal:2020wfl,Akal:2020twv,Numasawa:2020sty,Colin-Ellerin:2020mva,Basak:2020aaa,Geng:2020fxl,Caceres:2020jcn,
Deng:2020ent,Choudhury:2020hil,Karananas:2020fwx,Bousso:2021sji,Verheijden:2021yrb,
Patrascu:2021fyg,May:2021zyu,Kawabata:2021hac,Anderson:2021vof,Bhattacharya:2021jrn,
Kim:2021gzd,Hollowood:2021nlo,Wang:2021mqq,Miyata:2021ncm,Ghosh:2021axl,
Balasubramanian:2021wgd,Uhlemann:2021nhu,Neuenfeld:2021bsb,Li:2021lfo,
Geng:2021iyq,Geng:2021wcq,Bachas:2021fqo,Wang:2021woy,Fallows:2021sge,
Aalsma:2021bit,Karlsson:2021vlh,Qi:2021sxb}.

It is surprising that a semi-classical formula such as quantum extremal surface can capture the unitarity of quantum gravity. In two dimensional Jackiw-Teitelboim (JT) gravity plus CFT model, the QES formula can be derived from the Euclidean gravitational path integral, or the so called replica wormhole calculation~\cite{Almheiri:2019qdq}. However there appears the factorization puzzle or JT/ensemble relation if one takes the replica wormhole solutions seriously~\cite{Penington:2019kki}. It is therefore interesting to ask how we can justify QES formula by other means. In particular, as a semi-classical formula at this stage, what would be the potential UV correction?

In~\cite{Deng:2020ent} we proposed defect extremal surface (DES) formula as the holographic counterpart for the boundary QES formula. Defect extremal surface is defined by minimizing the Ryu-Takayanagi surface corrected by the defect theory. That is particularly interesting when the RT surface crosses or terminates on the defect. In a static set up of AdS$_3$/BCFT$_2$, it was demonstrated that the defect extremal surface formula gives precisely the same entanglement entropy as that from the boundary quantum extremal surface. In particular a decomposition procedure consist of partial Randall-Sundrum reduction and Maldacena duality has been proposed, for an AdS bulk with a defect brane, from which one can clearly see how island formula emerges from a brane world system with gravity glued to a flat space quantum field theory.

In this paper we extend our study on defect extremal surface to time dependent cases as well as higher dimensions. The main motivation to consider the time dependent defect extremal surface is to derive Page curve. We start from a Euclidean AdS$_3$/BCFT$_2$ and conformally transform the boundary (including the brane) to a cylinder bath with a Euclidean time circle (temperature) glued with a brane. We then cut off half of the system and construct an initial state by defining an Euclidean path integral over the remaining half. This is essentially a Thermal Field Double state and we then evolve it along real time. In real time, one can see an induced eternal black hole on the brane. From boundary point of view, this is very similar to the system consist of an eternal black hole plus CFT bath discussed in~\cite{Almheiri:2019qdq}. Also the brane approach to this system was first explored by Rozali, Sully, Raamsdonk, Waddell and Wakeham in~\cite{Rozali:2019day}. However, compared with those works, there are several major differences here: First, we do not have JT gravity in the brane region. Rather we consider the $2d$ gravity on the brane purely from the partial reduction of the bulk. Second, we do have quantum field theory on the brane, which is considered as the defect theory since we treat brane as a defect in AdS. In particular, compared with the holographic set up~\cite{Rozali:2019day}, we obtain a $2d$ effective description following the decomposition procedure consist of partial Randall-Sundrum reduction and Maldacena duality proposed in~\cite{Deng:2020ent}.

We first derive Page curve from defect extremal surface formula and then compute it independently using island formula in the $2d$ effective description mentioned above. We find precise agreement. This justifies the validity of defect extremal surface formula in time dependent set up. We then move to higher dimensions. There is no simple tools to calculate the matter entanglement entropy of several intervals in higher dimensions. We therefore assume that our matter CFTs are holographic and employ the corresponding Ryu-Takayanagi results. The defect extremal surface calculation for the entanglement entropy in higher dimensions is then straightforward. To find the one-dimension lower description of the AdS$_{d+1}$ with a $d$-dimensional brane, we again employ the partial Randall-Sundrum reduction, which leads to a $d$-dimensional Newton constant on the brane (as the leading term in effective action). After that we perform an independent computation of the same entropy, now using QES formula in the $d$-dimensional effective description. Through this work we assume that brane CFT has the same central charge as bath CFT. We find some discrepancy between DES result and QES result, namely DES always gives smaller entropy. Our results indicate that the UV completion of island formula may give a smaller entropy in higher dimensions.

This paper is organized as follows. We review defect extremal surface in Section 2. After that we discuss AdS$_3$/BCFT$_2$ in a dynamical set up and derive Page curve in Section 3 from both DES approach and QES approach and find agreement. In Section 4, we move to higher dimensions and discuss entanglement entropy for a strip both from DES calculation and QES calculation, where we extend the decomposition of partial Randall-Sundrum+AdS/CFT to higher dimensions. In Section 5, we construct a higher dimensional eternal black hole on a brane following the set up in Section 3 and find the Page curve. We check the entanglement entropy for a ball shape subregion in higher dimensions in Section 6, as another example of higher dimensional DES and QES calculation. We found similar results as those in Section 4. We conclude and discuss future questions in Section 7.

\section{Review of defect extremal surface}\label{sec2}
In this section, we give a brief review of defect extremal surface. We first review some basics of AdS/BCFT and then bring in the DES proposal. Finally, we review how to obtain boundary effective description of the bulk and show that the entropy computed by DES in the bulk equals to the entropy computed by island formula in the boundary description~\cite{Deng:2020ent}.
\subsection{AdS/BCFT}
When there is a codimension one end of the world brane in AdS bulk, the total action is given by:
\bal\begin{split}
I&=\frac{1}{16 \pi G_{N}} \int_{N} \sqrt{-g}(R-2 \Lambda)+\frac{1}{8 \pi G_{N}} \int_{M} \sqrt{-\gamma}\left(K^{(\gamma)}-\Sigma^{(\gamma)}\right)\\
&+\frac{1}{8 \pi G_{N}} \int_{Q} \sqrt{-h}K^{(h)}+I_{Q}+I_{P}\ ,
\end{split}\eal
where $N$ denotes the bulk AdS spacetime, $M$ denotes the asymptotic boundary where the Dirichlet boundary condition is imposed. $Q$ is the brane where Neumann boundary condition is imposed. $I_{Q}$ is the matter action on $Q$ and $I_{P}$ is the counter term on $P=Q\cap M$. The variation of this action leads to the Neumann boundary condition on $Q$~\cite{Takayanagi:2011zk}
\bal\label{NBC}
K_{a b}^{(h)}-h_{a b} K^{(h)}=8\pi G_{N}T_{ab}\ ,
\eal
where $h_{a b}$ is the induced metric on the brane and $K_{a b}^{(h)}$ is the extrinsic curvature. $T_{ab}=-\frac{2}{\sqrt{-h}} \frac{\delta I_{Q}}{\delta h^{a b}}$ is the stress energy tensor coming from the variation of matter action.
Consider a brane with a constant tension, $I_Q=-\frac{1}{8 \pi G_{N}} \int_{Q} \sqrt{-h}T$, where $T$ denotes the brane tension, Neumann boundary condition is reduced to
\bal
K^{(h)}_{a b}=(K^{(h)}-T) h_{a b}.
\eal

\begin{figure}[h]\label{16}
  \centering
  \includegraphics[width=12cm,height=6cm]{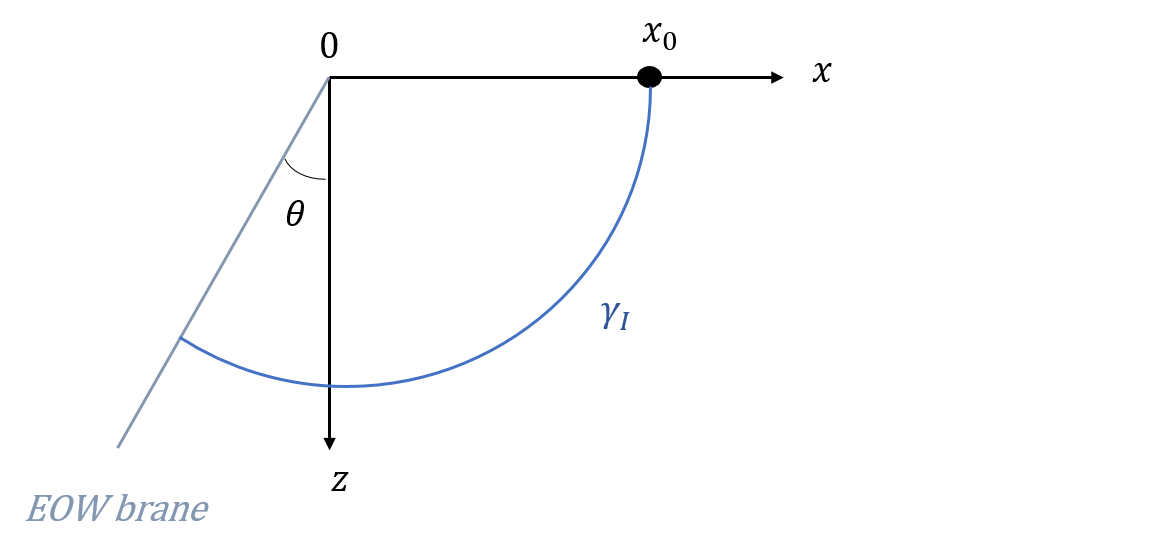}\\
  \caption{RT surface for an interval $I:=[0,x_0]$ that contains the boundary.}\label{16}
\end{figure}

In AdS$_3$/BCFT$_2$, the bulk is 3 dimensional and the $Q$ brane is 2 dimensional as shown in~Fig.\ref{16}. There are two sets of coordinates that are useful: $(t,x,z)$ and $(t,\rho,y)$. They are related by
\bal
\label{zyxy}
z=-y / \cosh \frac{\rho}{l}\ , \quad x=y \tanh \frac{\rho}{l}\,
\eal
and the bulk metric can be written as
\bal\label{2dmet}\begin{split}
d s^{2}_N&=d \rho^{2}+l^2\cosh ^{2} \frac{\rho}{l} \cdot \frac{-d t^{2}+d y^{2}}{y^{2}}\\
&=\frac{l^2}{z^2}(-dt^2+dz^2+dx^2),
\end{split}\eal
where $l$ is the AdS radius. It is also useful to introduce polar coordinate $\theta$ with $\frac{1}{\cos (\theta)}=\cosh \left(\frac{\rho}{l}\right)$.
If the brane locates at $\rho=\rho_0$, where $\rho_0$ is a positive constant, then the geometry on the brane is $\text{AdS}_2$ and the relation between $\rho_0$ and brane tension is $T=\frac{\tanh \left(\frac{\rho_0}{l}\right)}{l}$.

For an interval $I:=[0,x_0]$ in BCFT, the entanglement entropy can be computed holographically using RT formula. As shown in Fig.\ref{16}, the minimal surface denoted by $\gamma_I$ terminates on a point on the brane which can be determined by extremization. The entanglement entropy is
\bal\begin{split}
S_{I}=\frac{\operatorname{Area}\left(\gamma_{I}\right)}{4 G_{N}}&=\frac{c}{6} \log \frac{2x_0}{\epsilon}+\frac{c}{6}\rho_0\\
&=\frac{c}{6} \log \frac{2x_0}{\epsilon}+\frac{c}{6} \operatorname{arctanh}(\sin \theta_0),
\end{split}\eal
where $c$ is the CFT central charge and $\epsilon$ is the UV cut off.

\subsection{Bulk defect extremal surface}
When there is quantum matter localized on the end of the world brane, one should take into account its contribution when calculating entanglement entropy. It is obvious that one should do so if we treat the brane as a defect in the bulk. In the work of~\cite{Deng:2020ent}, it has been proposed that the entanglement entropy including defect contribution is given by defect extremal surface (DES) formula,
\bal\label{DES}
S_{\mathrm{DES}}=\min _{\Gamma, X}\left\{\operatorname{ext}_{\Gamma, X}\left[\frac{\operatorname{Area}(\Gamma)}{4 G_{N}}+S_{\text {defect }}[D]\right]\right\}, \quad X=\Gamma \cap D,
\eal
where $\Gamma$ is co-dimension two surface in AdS bulk and $X$ is the lower dimensional entangling surface given by the intersection of $\Gamma$ and the defect $D$.

As an example to illustrate, one can include CFT matter on the brane in the previous AdS$_3$/BCFT$_2$. The bulk action (disregarding $M$) is then given by
\bal
I=\frac{1}{16 \pi G_{N}} \int_{N} \sqrt{-g}(R-2 \Lambda)+\frac{1}{8 \pi G_{N}} \int_{Q} \sqrt{-h} (K-T) + I_{CFT}\ ,
\eal
and the vacuum one point function of the CFT stress tensor is given by
\bal
\label{st}
\langle T_{ab}\rangle_{\text{AdS}_2}=\chi h_{ab}.
\eal
The Neumann boundary condition then becomes
\bal
K_{a b}-h_{a b} (K-T)=8\pi G_{N}\chi h_{ab}\ .
\eal
For a brane located at constant $\rho_0$, the metric on the brane is again $\text{AdS}_2$. The entanglement entropy for an interval $I:=[0,x_0]$ can be calculated by DES formula (\ref{DES}).
The final result is
\bal\label{bgen}
S_{\text{DES}}=\frac{c}{6} \log \frac{2x_0}{\epsilon}+\frac{c}{6} \operatorname{arctanh}(\sin \theta_0)+\frac{c'}{6} \log \left(\frac{2l}{\epsilon_y \cos\theta_0}\right)\ ,
\eal
where $\theta_0$ is related to $\rho_0$ by $\frac{1}{\cos (\theta_0)}=\cosh \left(\frac{\rho_0}{l}\right)$ and $c'$ is the CFT central charge on the brane. Notice that in this case, the defect contribution on the AdS$_2$ brane is a constant, $S_{\text {defect }}=\frac{c'}{6} \log \left(\frac{2l}{\epsilon_y \cos\theta_0}\right)$, therefore it does not shift the position of the Ryu-Takayanagi surface.
\subsection{Boundary quantum extremal surface}
One can justify the above DES formula by island formula \cite{Almheiri:2019hni}. This has been carried out in~\cite{Deng:2020ent} by proposing an effective 2d description for the AdS$_3$ bulk with a brane. The effective 2d description was obtained by combining {\it partial Randall-Sundrum reduction}~\footnote{For the detail discussion of partial Randall-Sundrum, see~\cite{Deng:2020ent}.} and AdS/CFT correspondence as follows.

\begin{figure}[h]
  \centering
  \includegraphics[width=10cm,height=6cm]{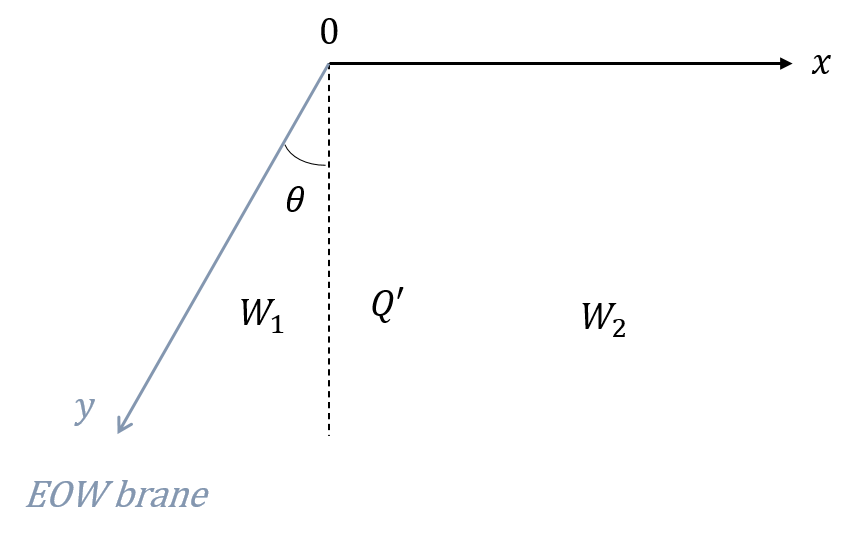}\\
  \caption{Bulk decomposition by inserting an imaginary boundary $Q'$.}\label{19}
\end{figure}

\begin{figure}[h]
  \centering
  \includegraphics[width=15cm,height=5cm]{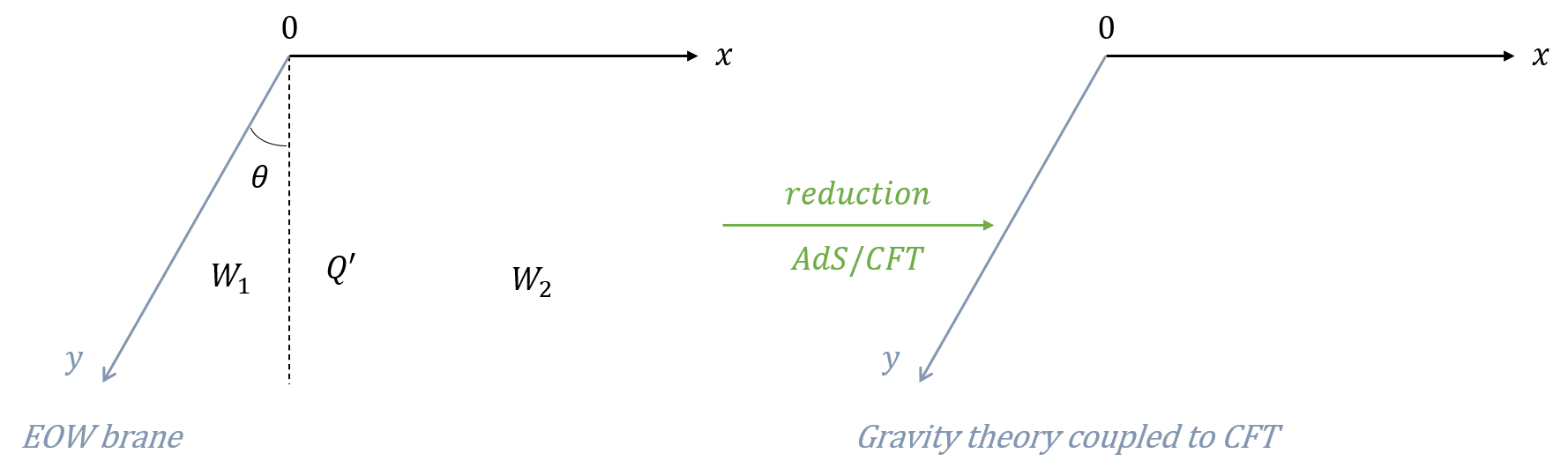}\\
  \caption{Effective description.}\label{20}
\end{figure}

We first insert an imaginary boundary $Q'$ that is orthogonal to the asymptotic boundary as shown in Fig.\ref{19}, and the bulk is decomposed into two parts $W_1$ and $W_2$. For $W_1$ we employ a partial Randall-Sundrum reduction along the extra dimension $\rho$, and the resulting brane theory is a 2d gravity coupled with CFT matter. According to AdS/CFT, $W_2$ can be dual to half space CFT with zero boundary entropy. In the end we have a gravity theory coupled with a CFT on the brane, glued with a flat space CFT as shown in Fig.\ref{20}. Notice that the boundary condition between 2d brane theory and half space CFT is transparent and the imaginary boundary $Q'$ is essentially the holographic dual of the transparent boundary condition.


One can use island formula to compute entanglement entropy in this 2d effective description of the system. As an illustration, consider an interval $[0,L]$ in the flat CFT region. According to island formula (quantum extremal surface), the entropy is calculated by
\bal
S_{\text{QES}}=\text{ext}_a\big\{S_{\text{gen}}(a)\big\}=\text{ext}_a\big\{S_{\text{area}}
(y=a)+S_{\text{matter}}([-a,L])\big\},
\eal
where $a$ is the boundary of the island on the brane.

The area term can be obtained as follows. When doing reduction along $\rho$ direction, $d+1$-dimensional gravity on the wedge is reduced to a $d$-dimensional gravity on the brane. The $d$-dimensional Newton constant on the brane is~\cite{Deng:2020ent}
\bal
\frac{1}{G_{N}^{(d)}}=\frac{1}{G_{N}}\left(\cosh \frac{\rho_0}{l}\right)^{2-d} \int_{0}^{\rho_0} d \rho\left(\cosh \frac{\rho}{l}\right)^{d-2}\ .
\eal
Notice that in 2 dimensions the area term is
\bal
S_{\text{area}}(y=a)=\frac{1}{4G_{N}^{(2)}}=\frac{\rho_0}{4G_{N}}=\frac{c}{6}\text{arctanh} (\sin \theta_0)\ .
\eal
After extremization, the final result of the entanglement entropy is
\bal\begin{split}
S_{\text{QES}}&=\frac{c}{6} \operatorname{arctanh}(\sin \theta_0)+\frac{c}{6}\log \frac{4x_0l}{\cos\theta_0\epsilon\epsilon_y}\\
&=\frac{c}{6} \log \frac{2x_0}{\epsilon}+\frac{c}{6} \operatorname{arctanh}(\sin \theta_0)+\frac{c}{6} \log \frac{2l}{\epsilon_y \cos\theta_0}\ .
\end{split}\eal
This is exactly the same entropy as calculated by DES for $c'=c$, which justified the DES proposal. Inversely, defect extremal surface together with partial Randall-Sundrum gives a holographic derivation of island formula. In AdS$_3$/BCFT$_2$, it is therefore clear that defect extremal surface formula is the holographic counterpart of boundary island formula. Our main goal in this paper is to study further defect extremal surfaces in time dependent cases as well as in higher dimensions.

\section{Page curve for $2d$ eternal black hole}\label{sec4}
In this section, we study the time dependent AdS$_3$/BCFT$_2$, where an eternal black hole emerges on the EOW brane. Following the decomposition procedure in the previous section, we obtain a 2d effective theory to describe the black hole evaporation.
We then compute the Page curve using island formula and find that it agrees with the bulk defect extremal surface result precisely. We compute in Euclidean spacetime and then analytically continue to real time. The holographic computation without considering defect contribution has been done in~\cite{Rozali:2019day}.
\subsection{The system}
\begin{figure}[h]
  \centering
  \includegraphics[width=13cm,height=8cm]{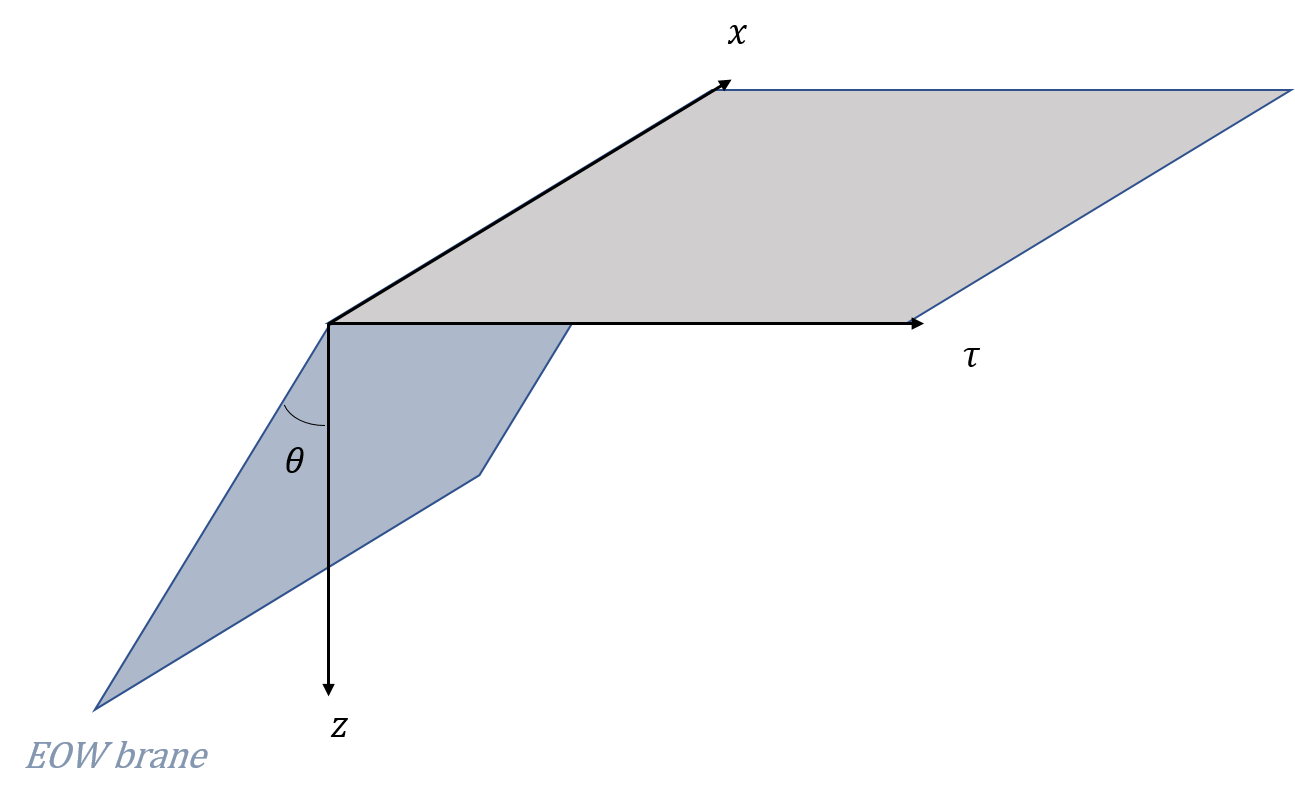}\\
  \caption{Holographic dual of Euclidean BCFT defined on half spacetime $(\tau>0)$. }\label{1}
\end{figure}

We first look at how the eternal black hole emerges. Recall that the holographic dual of the BCFT defined on a half spacetime is given by an AdS with an EOW brane. In Euclidean spacetime, there is no difference between space and time if we do not try to give physical interpretation, one can therefore choose the boundary of BCFT to be $\tau=0$ and the BCFT is defined in the region $\tau \ge 0$. The metric of the AdS is given by
\begin{equation}
\begin{split}
ds^2=l^2\frac{d\tau^2+dx^2+dz^2}{z^2}\ ,
\end{split}
\end{equation}
with the holographic region $\tau+z\tan \theta>0$. As shown in Fig.\ref{1}, the EOW brane in the AdS is located at $\tau=-z\tan \theta$.

\begin{figure}[h]
  \centering
  \includegraphics[width=13cm,height=8cm]{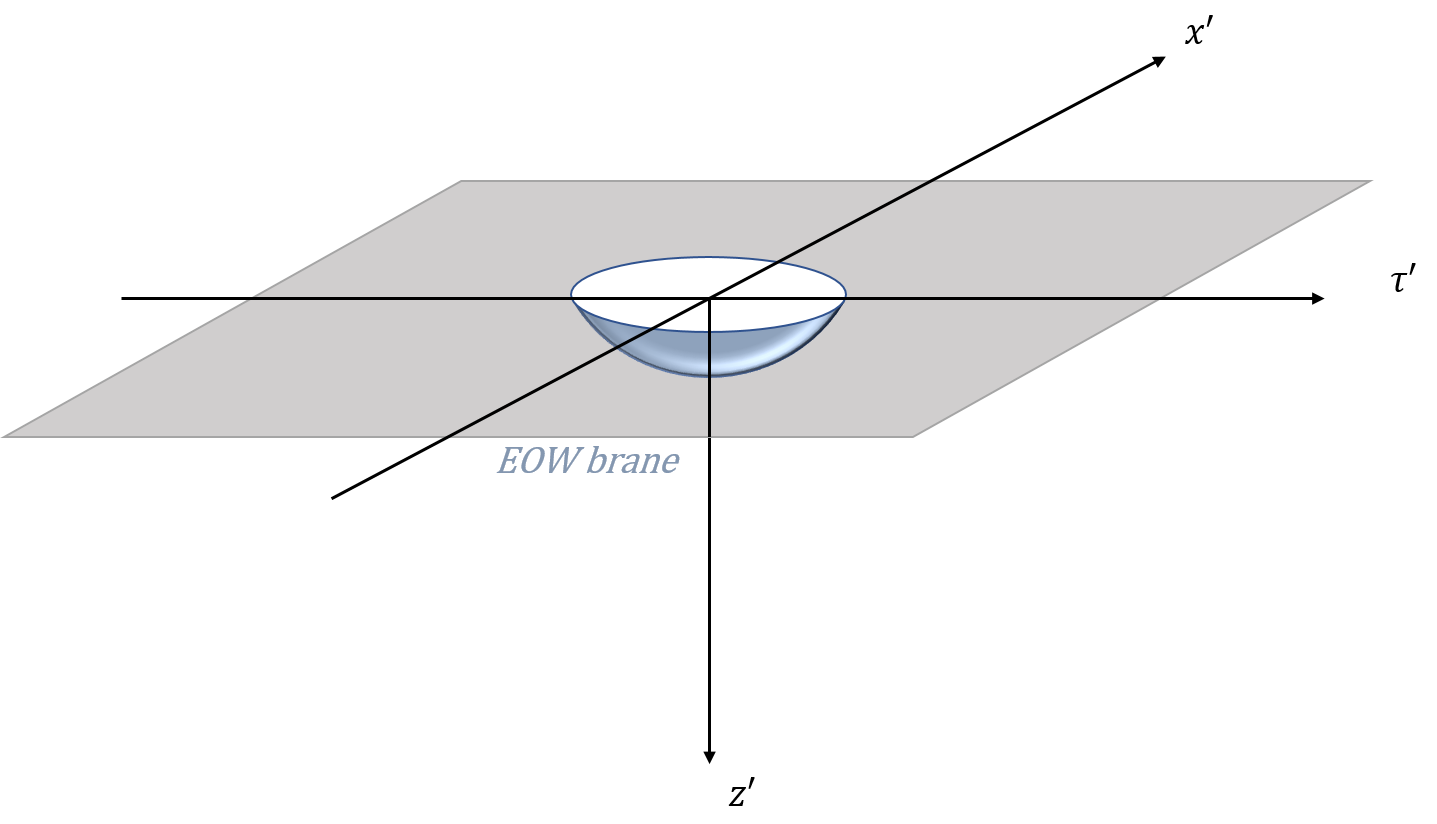}\\
  \caption{Holographic dual of a BCFT after the conformal transformation. }\label{2}
\end{figure}

Under particular conformal transformations
\begin{equation}
\label{sct}
\begin{split}
\tau&=\frac{2(x'^2+\tau'^2+z'^2-1)}{(\tau'+1)^2+x'^2+z'^2}\ ,\\
x&=\frac{4x'}{(\tau'+1)^2+x'^2+z'^2}\ ,\\
z&=\frac{4z'}{(\tau'+1)^2+x'^2+z'^2}\ ,
\end{split}
\end{equation}
the boundary is mapped to a circle $$x'^2+\tau'^2=1$$ and the EOW brane is mapped to a part of sphere $$(z'+\tan\theta)^2+x'^2+\tau'^2=\sec^2\theta\ ,$$ while the metric is preserved, as shown in Fig.\ref{2}. Assuming the UV cut off where the BCFT lives is $z'=\epsilon$, it will not be a constant in the original coordinate system $(\tau,x,z)$ as seen from the last terms of (\ref{cDES})(\ref{RT1}).

\begin{figure}
  \centering
  \includegraphics[width=6cm,height=6cm]{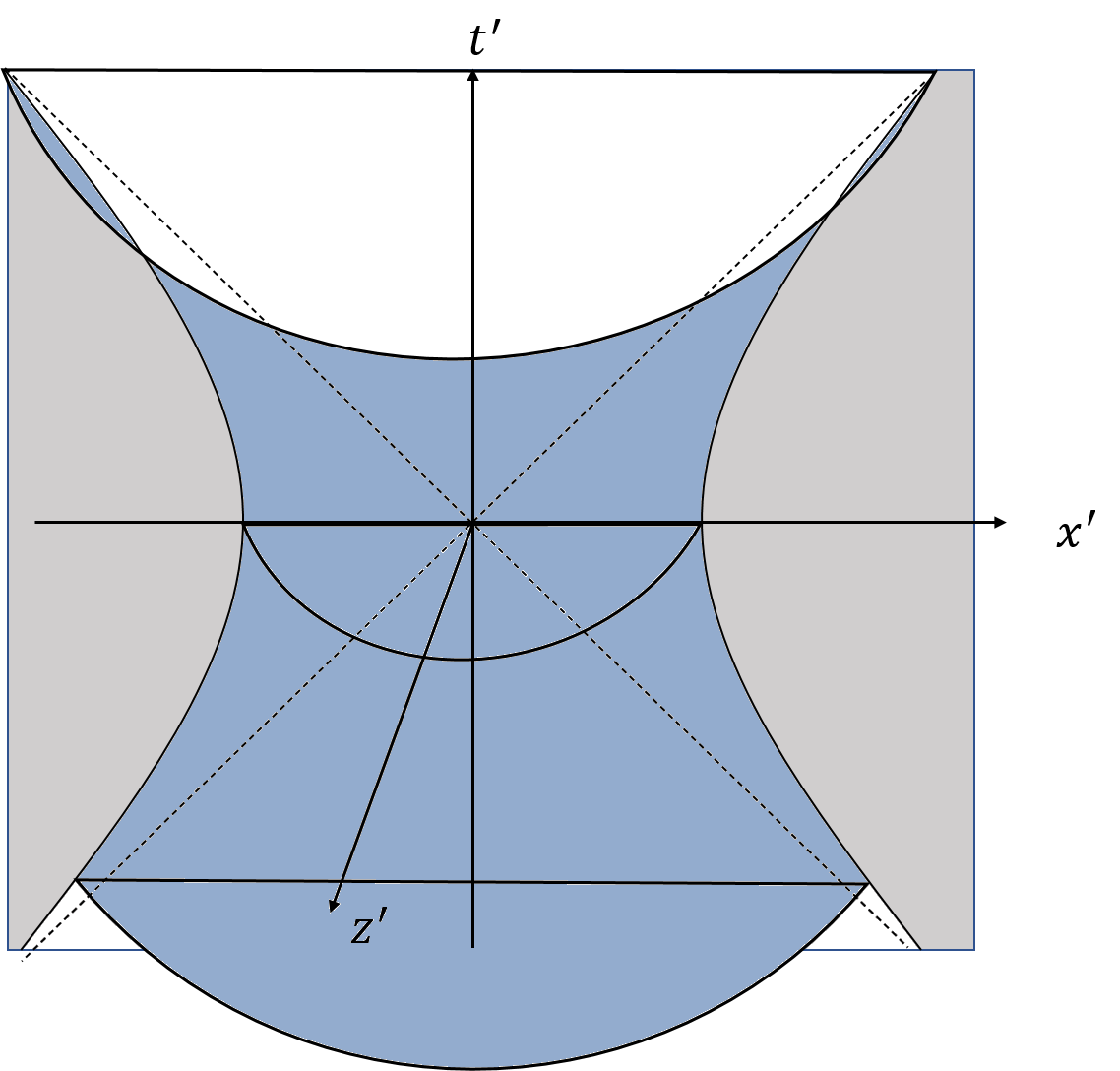}\\
  \caption{The horizon on the brane. }\label{3}
\end{figure}

By doing wick rotation $\tau'\rightarrow it'$, one can find that the light-like curves on the EOW brane
\begin{equation}
x'=\pm t'\ ,\quad z'={1-\sin\theta\over \cos\theta}
\end{equation} asymptotes to the boundary of the brane~\cite{Rozali:2019day}, $x'^2-t'^2=1$, when $t'\to \infty$. These are actually the black hole horizons and the black hole interior is given by $|x|<t$ or $z>{1-\sin\theta\over \cos\theta}$.
In Fig.\ref{3} the horizon on the brane is projected to be dashed lines on the asymptotic boundary.

\begin{figure}
  \centering
  \includegraphics[width=10cm,height=4cm]{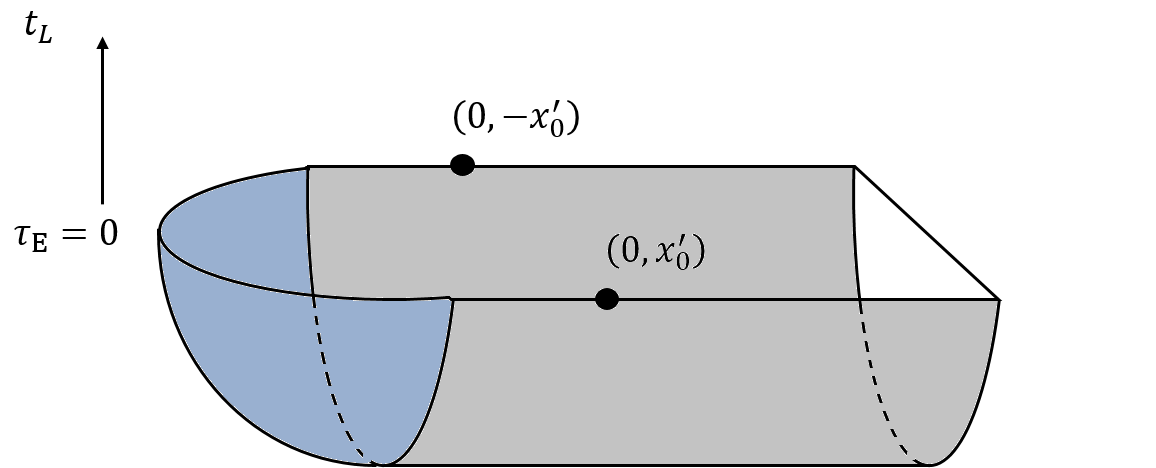}\\
  \caption{Thermal field double perspective of the system.}\label{12}
\end{figure}

Following the decomposition procedure in the previous section, we will eventually obtain a gravity system on the brane glued to a bath. The angle direction in polar coordinates of $x'-\tau'$ plane is naturally identified as Euclidean time circle. To factorize the time circle we need another coordinate transformation
\begin{equation}
x'= e^{X}\cos\phi\ ,\quad \tau'=e^{X}\sin\phi\ ,
\end{equation}
which sends the bath CFT onto a cylinder. The Euclidean path integral on half of the cylinder essentially prepares the initial state as a TFD state as shown in Fig. \ref{12}. Finally we Wick rotate $\phi$ to real time $T$ to see the nontrivial evolution, which can be equivalently written as
\begin{equation}
\begin{split}
x'=e^X\cosh T,\ \quad\tau'=ie^X\sinh T\ .
\end{split}
\end{equation}

\subsection{Bulk DES}\label{}
\begin{figure}[h]
  \centering
  \includegraphics[width=13cm,height=8cm]{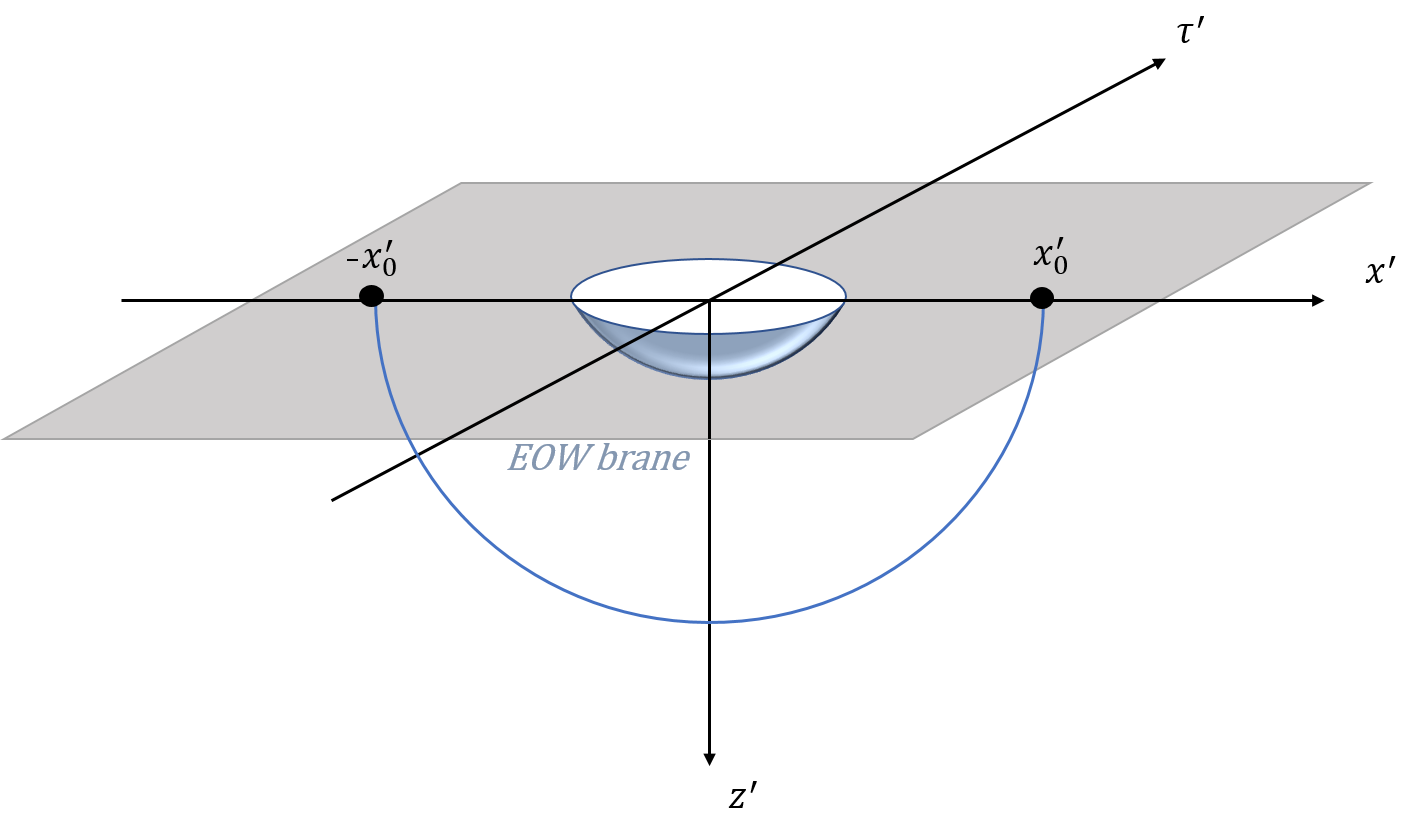}\\
  \caption{Connected phase of extremal surface.}\label{5}
\end{figure}

We consider the interval $[-\infty,-x'_0]\cup[x'_0,\infty]$ at $\tau'=\tau'_0$ and use the DES formula to calculate the entanglement entropy. The two endpoints $(\tau'_0,x'_0)$ and $(\tau'_0,-x'_0)$ are mapped to $(\tau_0,x_0)$ and $(\tau_0,-x_0)$ respectively by (\ref{sct}). There are two phases of the extremal surface, one is connected and the other is disconnected. In the former phase, the extremal surface does not intersect with the EOW brane as shown in Fig.\ref{5}. Therefore, no bulk term would be included, and the entropy is given by the RT surface,
 \begin{equation}
 \label{cDES}
\begin{split}
S_{\text{DES}}=&\frac{c}{6}\bigg[\log (2x_0)^2-2\log \frac{4\epsilon}{(\tau'_0+1)^2+x'^2_0}\bigg]\\
=&\frac{c}{3}\log \frac{2 x'_0}{\epsilon}\ .
\end{split}
\end{equation}

\begin{figure}[h]
  \centering
  \includegraphics[width=13cm,height=8cm]{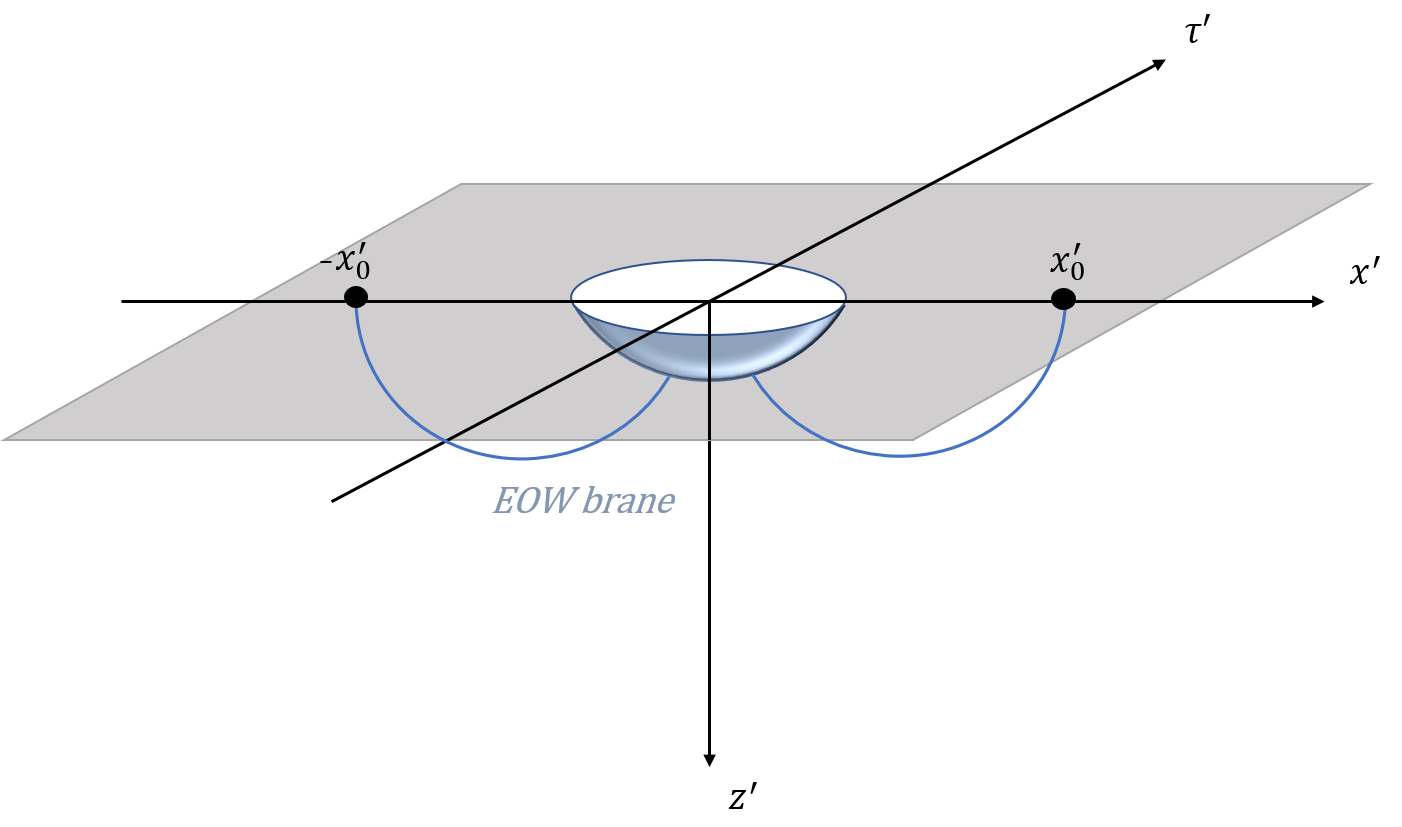}\\
  \caption{Disconnected phase of extremal surface.}\label{6}
\end{figure}

In the disconnected phase, the extremal surfaces intersect with the brane at two points as shown in Fig.\ref{6}. By the symmetry with respect to $x=0$ plane, the locations of the two intersection points can be denoted as $(\tau_1',\pm x_1',z_1')$, or $(-z_1\tan\theta,\pm x_1,z_1)$ in the coordinate system $(\tau ,x,z)$. The length of each extremal surface is given by
\begin{equation}
\label{RT1}
\begin{split}
\frac{4G_N}{l}S_{\text{RT}_1}=&\frac{4G_N}{l}S_{\text{RT}_2}\\
=&\log \frac{(\tau_0+z_1\tan\theta)^2+(x_0-x_1)^2+z_1^2}{\sqrt{(\tau_0+z_1\tan\theta)^2+(x_0-x_1)^2}}\\
&+\text{arctanh}\frac{(\tau_0+z_1\tan\theta)^2+(x_0-x_1)^2-z_1^2}{(\tau_0+z_1\tan\theta)^2+(x_0-x_1)^2+z_1^2}\\
&-\log \frac{4\epsilon}{(\tau'_0+1)^2+x'^2_0}\ ,
\end{split}
\end{equation}
where the last term corresponds to the cut off in the coordinates without prime.

To compute the defect contribution, i.e. the entropy of the interval bounded by the intersection points on the brane, we insert two twist operators. From the correlation function of the twist operators, we get the entropy as follows,
\begin{equation}
\label{btAB}
\begin{split}
S_{\text{defect}}([A,B])&=\lim_{n\to 1}\frac{1}{1-n}\log\langle \Psi_n(A)\bar{\Psi}_n(B)\rangle_Q\\
&=\frac{c}{3}\min\left\{\log\frac{2lx_1}{\epsilon_yz_1},\log\frac{2l}{\epsilon_y\cos\theta} \right\}\ .
\end{split}
\end{equation}
By combining the area terms (\ref{RT1}) with the bulk term (\ref{btAB}), we have the generalized entropy
\begin{equation}
\label{genAB}
\begin{split}
S_{\text{gen}}([A,B])=S_{\text{RT}_1}+S_{\text{RT}_2}+S_{\text{defect}}([A,B])\ .
\end{split}
\end{equation}

If $x_1\cos \theta<z_1$, the first choice in (\ref{btAB}) is picked. However, it turns out that $\frac{\partial S_{\text{gen}}}{\partial x_1}$ as well as $\frac{\partial S_{\text{gen}}}{\partial z_1}$ never vanishes. In other words, there is no DES solution. When $x_1\cos \theta>z_1$, in which case the second choice in (\ref{btAB}) is picked, we find that DES solution coincides with the RT surface since the bulk term is constant. More specifically, the DES solution is
\bal
\begin{split}
\begin{cases}
z_1=\tau_0\cos\theta\\
x_1=x_0\ .
\end{cases}
\end{split}
\eal
And the restriction $x_1\cos \theta>z_1$ becomes $x_0>\tau_0$, or
\begin{equation}
\label{res1}
\begin{split}
2x'_0>x'^2_0+\tau'^2_0-1
\end{split}
\end{equation}
in the coordinate system $(\tau',x')$. From the extremal condition we calculate that the entropy is
\begin{equation}
\label{dDES}
\begin{split}
S_{\text{DES}}=\frac{c}{3}\left(\log \frac{x'^2_0+\tau'^2_0-1}{\epsilon}+\text{arctanh}\sin \theta+ \log\frac{2l}{\epsilon_y\cos\theta}  \right)\ .
\end{split}
\end{equation}
Comparing (\ref{cDES}) and (\ref{dDES}), we can find that when the later is favored,
\begin{equation}
\begin{split}
2x'_0>(x'^2_0+\tau'^2_0-1)e^{\text{arctanh}(\sin \theta)}\frac{2l}{\epsilon_y \cos\theta}\ .
\end{split}
\end{equation}
Note that it is stronger than the restriction (\ref{res1}). Thus in the disconnected phase the second choice in (\ref{btAB}) is favored and the extremal point does exist.

\begin{figure}[htbp]
	\centering
	\includegraphics[width=9cm]{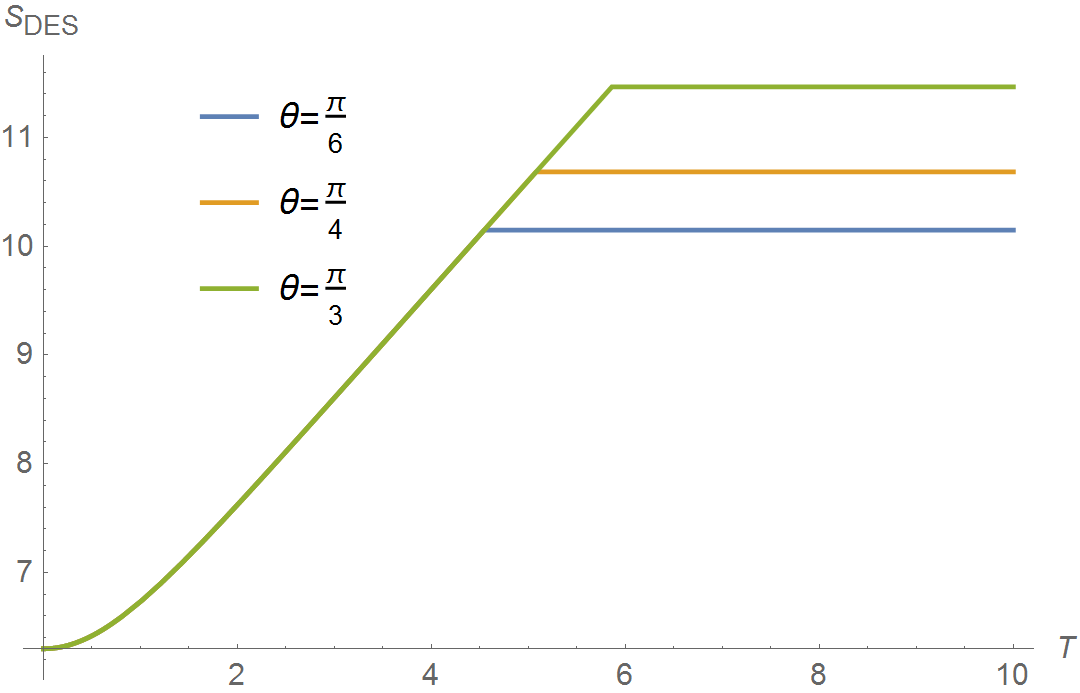}\\
	\caption{The entropy $S_\text{DES}$ (in the unit of $\frac{c}{3}$) with respect to time $T$ for $X_0=1$ and $\theta=\frac{\pi}{6},\frac{\pi}{4},\frac{\pi}{3}$. We pick $\epsilon=0.01$, $\epsilon_y=0.1$ and $l=1$.}
	\label{2dpc}
	\end{figure}

We summarize the result by writing the entropy in the coordinate system $(T,X)$
\begin{equation}
\label{tden}
\begin{split}
S_{\text{DES}}=\begin{cases}\frac{c}{3}\left(\log \frac{2\cosh T}{\epsilon}+X_0\right),\quad &T<T_P\\
\frac{c}{3}\left(\log \frac{e^{2X_0}-1}{\epsilon}+\text{arctanh}(\sin\theta)+\log\frac{2l}{\epsilon_y\cos\theta}\right),\quad &T>T_P\ .
\end{cases}
\end{split}
\end{equation}
This can be interpreted as the Von Neumann entropy of the bath as a function of physical time $T$, which fits the Page curve.
The Page time is at
\begin{equation}
\label{pt}
\begin{split}
T_P=\text{arccosh}\left( \sinh X_0 e^{\text{arctanh}\sin \theta}\frac{2l}{\epsilon_y \cos\theta}\right)\ .
\end{split}
\end{equation}
As explicitly shown in Fig.\ref{2dpc}, the entropy follows a Page curve increasing at early time and being constant after the Page time (\ref{pt}). We can also see that the Page time is larger for larger angle $\theta$ of the brane.
\subsection{Boundary QES}\label{}
Now we rederive the entropy of the interval $[-\infty,-x'_0]\cup[x'_0,\infty]$ at $\tau'=\tau'_0$ from the boundary point of view following the decomposition procedure in the previous section. Similar to DES, there are two possible phases in the $2d$ QES computation, one of which contains no contribution from the brane while the other includes the area term as well as the matter entropy from the brane.

Without contribution from the brane, the entropy of $[-\infty,-x'_0]\cup[x'_0,\infty]$ is just the matter entropy, i.e.
\bal\label{nisp}
S_{\text{QES}}=S_{\text{matter}}([-\infty,-x'_0]\cup[x'_0,\infty])=\frac{c}{3}\log \frac{2x'_0}{\epsilon}\ ,
\eal
which is the same as (\ref{cDES}).

Since the brane CFT is coupled to gravity, there is also a possibility that the matter term receives an interval contribution on the brane with the endpoints at $A:(y_1,x_1)$ and $B:(y_1,-x_1)$. These two endpoints also bring area terms, i.e.
\bal
\label{at}
S_{\text{area}}=2\times \frac{1}{4G_N^{(2)}}=\frac{c}{3}\text{arctanh}\sin\theta\ .
\eal
By employing the entropy formula of two disjoint intervals at large central charge~\cite{Hartman:2013mia}, we get the matter term as
\bal
\begin{split}
\label{mtQES}
S_{\text{matter}}\big(A,B\big)=&\frac{c}{3}\min \biggr\{ \log \frac{2x_1x_0l}{y_1\cos \theta  \epsilon_y}, \log \frac{\left[(y_1+\tau_0)^2+(x_1-x_0)^2\right]l}{y_1\cos \theta \epsilon_y} \biggr\}\\
&-\frac{c}{3}\log \frac{4\epsilon}{(\tau'_0+1)^2+x'^2_0}\ .
\end{split}
\eal
Notice that the last term corresponds to the cut-offs at the endpoints $(\tau_0,x_0)$ and $(\tau_0,-x_0)$. Now, if the first choice in the ``min" is picked, the extremal condition $\partial_{x_1}S_{gen}(A,B)=\frac{c}{3}\frac{1}{x_1}=0$ as well as $\partial_{y_1}S_{gen}(A,B)=-\frac{c}{3}\frac{1}{y_1}=0$ has no solution. If the second choice is picked, we can find that the extremization procedure gives
\bal
\begin{split}
\begin{cases}y_1=\tau_0\\
x_1=x_0\ .
\end{cases}
\end{split}
\eal
Combined with the area term (\ref{at}), it thus gives the QES result of the entropy,
\bal
\label{isp}
S_{\text{QES}}=\frac{c}{3}\text{arctanh}\sin\theta+ \frac{c}{3}\log \frac{2({x'_0}^2+{\tau'_0}^2-1)l}{\cos \theta\epsilon \epsilon_y}\ .
\eal
To summarize in $(X, T)$ coordinates,
\begin{equation}
\begin{split}
S_{\text{QES}}=\begin{cases}\frac{c}{3}\left(\log \frac{2\cosh T}{\epsilon}+X_0\right),\quad &T<T_P\\
\frac{c}{3}\left(\log \frac{e^{2X_0}-1}{\epsilon}+\text{arctanh}\sin\theta+\log\frac{2l}{\epsilon_y\cos\theta}\right),\quad &T>T_P\ ,
\end{cases}
\end{split}
\end{equation}
which is exactly the same as (\ref{tden}). Similarly, it can also be checked that the second choice in (\ref{mtQES}) is indeed the minimum after the Page time.
\section{Entanglement entropy for a strip in BCFT$_d$}\label{}
From now on, we study defect extremal surface in higher dimensions.
We first look at the entanglement entropy for a strip in half-space BCFT$_d$ defined on $x>0$.~\footnote{This BCFT should be distinguished from that in the previous section since it is static.} The left boundary of the strip is at $(x=0,\tau_1)$ and the right boundary $A$ is at $(x_1,\tau_1)$. The holographic dual of the BCFT is given by AdS$_{d+1}$ with an EOW brane. The metric describing an AdS$_{d+1}$ is given by
\begin{equation}
\begin{split}
ds^2=l^2\frac{d\tau^2+dx^2+dz^2+d\vec{u}^2(=du_1^2+\cdots +du_{d-2}^2)}{z^2}\ ,
\end{split}
\end{equation}
with the bulk region given by $x+z\tan \theta>0$. The EOW brane in the AdS is located at $x=-z\tan \theta$. To proceed the calculation, we also cut off the strip in the $u_i$ directions, more explicitly, $-L/2<u_i<L/2,i=1,2,\cdots,d-2$. The strip is shown in Fig.\ref{17}.
\begin{figure}[h]\label{17}
  \centering
  \includegraphics[width=13cm,height=7.5cm]{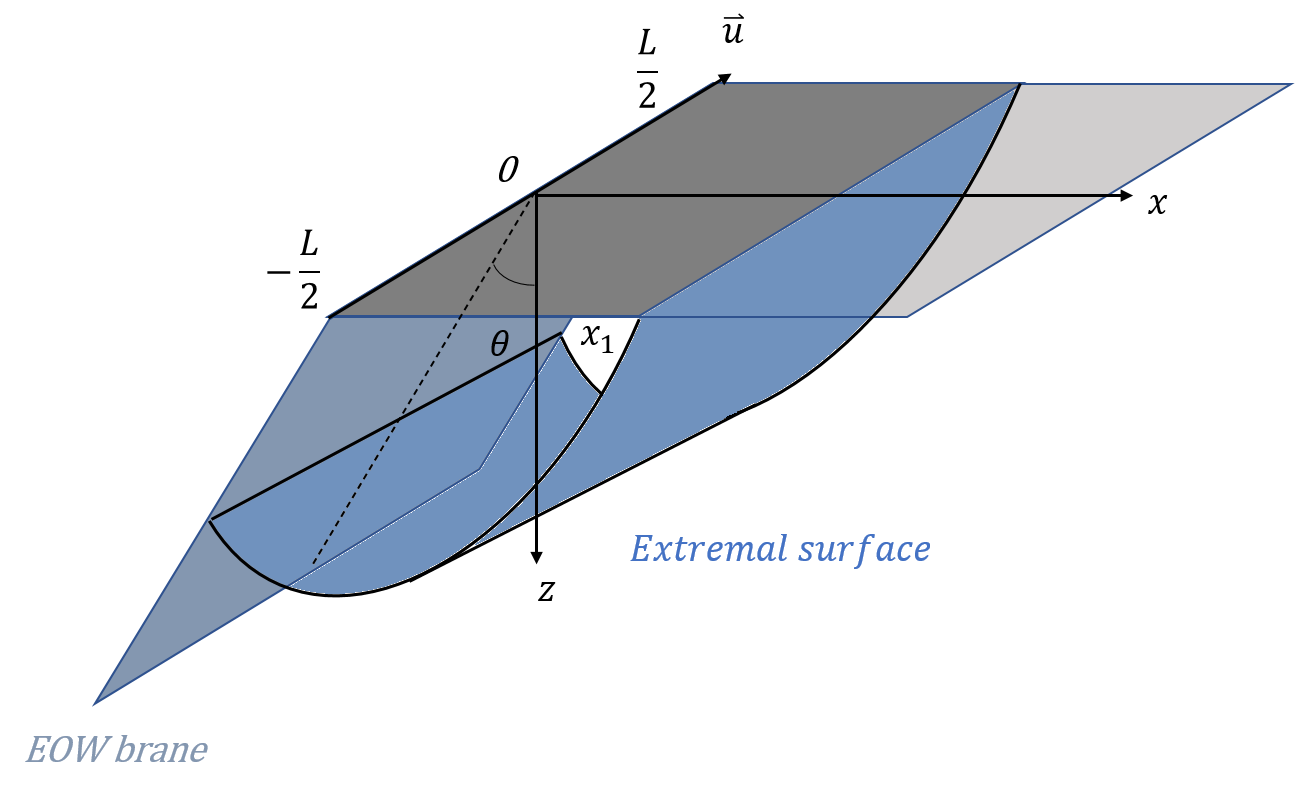}\\
  \caption{Strip and the extremal surface in BCFT.}\label{17}
\end{figure}
\subsection{Bulk DES}\label{}
We use the DES formula to calculate the entropy. It consists of an area term $S_{\text{area}}$ of the extremal surface and a bulk term $S_{\text{defect}}$ contributed by the entropy on the brane. In the present case with a translation symmetry in the direction of $u_i,i=1,2,\cdots, d-2$, the region for $S_{\text{defect}}$ is also a strip with the boundaries at $(y_a,\tau_a)$ (recall that the coordinate $y$ is related to $z$ and $x$ as (\ref{zyxy})), which is connected with $A$ by the extremal surface. For the static system, we have $\tau_a=\tau_1$. We can then write the generalized entropy as a function of $y_a$, i.e.
\begin{equation}
\begin{split}
S_{\text{gen}}(y_a)=S_{\text{RT}}(y_a)+S_{\text{defect}}(y_a)\ .
\end{split}
\end{equation}
For the first term, we have~\cite{Ryu:2006ef}
\begin{equation}
\label{RTa}
\begin{split}
4G_N^{(d+1)}S_{\text{RT}}(y_a)=&\frac{l^{d-1}}{d-2}\left(\frac{L}{\epsilon}\right)^{d-2}+\frac{\sqrt{\pi}}{2d-2}\frac{\Gamma(\frac{2-d}{2d-2})}{\Gamma(\frac{1}{2d-2})}l^{d-1}\left(\frac{L}{z_*}\right)^{d-2}\\
&+\eta l^{d-1}L^{d-2}\int_{y_a\cos \theta}^{z_*}dz\frac{z_*^{d-1}}{z^{d-1}\sqrt{z_*^{2d-2}-z^{2d-2}}}\ ,
\end{split}
\end{equation}
where $z_*$ as a function of $y_a$ is the turning point of the RT surface (or its extension) and $\eta=\pm 1$ depends on whether the turning point is on the RT surface or on the extension. Note that the first two terms correspond to half of the full RT surface ending on the asymptotic boundary and the last term is for the rest part. The integral turns out to be
\begin{equation}
\begin{split}
\frac{1}{z_*^{d-2}(d-2)}\bigg[&\left(\frac{y_a\cos \theta}{z_*}\right)^{2-d}F\left(\frac{1}{2},-\frac{d-2}{2(d-1)};\frac{d}{2(d-1)};\left(\frac{y_a\cos \theta}{z_*}\right)^{2d-2}\right)\\
&-F\left(\frac{1}{2},-\frac{d-2}{2(d-1)};\frac{d}{2(d-1)};1\right) \bigg]\ ,
\end{split}
\end{equation}
where $F$ denotes the hypergeometric function $_2F_1$ for short. The relation between $y_a$ and $z_*$ can be found by integrating the differential equation for the RT surface as follows.
\begin{equation}
\label{dzdx}
\begin{split}
&\frac{dz}{dx}=\frac{\sqrt{z_*^{2d-2}-z^{2d-2}}}{z^{d-1}}\\
\Rightarrow & \int_{y_a \cos \theta}^{z_*} dz \frac{z^{d-1}}{\sqrt{z_*^{2d-2}-z^{2d-2}}}=|x(z_*)+y_a\sin \theta |\\
\Rightarrow &-\frac{z_*}{d}\bigg[ \left(\frac{y_a\cos \theta}{z_*}\right)^dF\left(\frac{1}{2},\frac{d}{2(d-1)};\frac{3d-2}{2(d-1)};\left(\frac{y_a\cos \theta}{z_*}\right)^{2(d-1)}\right) \\
&-F\left(\frac{1}{2},\frac{d}{2(d-1)};\frac{3d-2}{2(d-1)};1\right)  \bigg]=\eta \left(x_1-\frac{\sqrt{\pi}\Gamma(\frac{d}{2d-2})}{\Gamma(\frac{1}{2d-2})}z_*+y_a\sin \theta \right)\ .
\end{split}
\end{equation}
From the above equation we can see that $\eta=\text{sign}(x_1-\frac{\sqrt{\pi}\Gamma(\frac{d}{2d-2})}{\Gamma(\frac{1}{2d-2})}z_*+y_a\sin \theta)$. Now we replace the variables $y_a$ and $z_*$ with $w_a\equiv \frac{y_a\cos \theta}{z_*}$ and $v_a\equiv \frac{1}{z_*}$. Then, the above equation simplifies to
\begin{equation}
\begin{split}
v_a=&-\frac{\eta}{x_1d}\bigg[ w_a^dF\left(\frac{1}{2},\frac{d}{2(d-1)};\frac{3d-2}{2(d-1)};w_a^{2(d-1)}\right)-F\left(\frac{1}{2},\frac{d}{2(d-1)};\frac{3d-2}{2(d-1)};1\right)  \bigg]\\
&+\frac{\sqrt{\pi}\Gamma(\frac{d}{2d-2})}{x_1\Gamma(\frac{1}{2d-2})}-\frac{w_a\tan \theta}{x_1}\ ,
\end{split}
\end{equation}
which gives $v_a$ as a function of $w_a$.

Now, we can calculate $S_{\text{defect}}$ in a similar way since the defect theory on the brane is also a BCFT, which has no boundary degree of freedom. Note that the background is curved and in order to bring this effect in,  one can calculate $S_{\text{defect}}$ holographically by replacing the flat space cut off $\epsilon_y$ with geodesic cut off $\Omega\epsilon_y$ where $\Omega$ is the conformal factor. By taking $\epsilon_y\rightarrow\Omega\epsilon_y=\frac{y_a\cos\theta\epsilon_y}{l}$, one can get
\bal\label{Sdef}
S_{\text{defect}}(y_a)=\frac{1}{4G_N^{(d+1)}}\bigg[\frac{l^{d-1}}{d-2}\left(\frac{Ll}{y_a
\cos\theta\epsilon_y}\right)^{d-2}+\frac{\pi^{\frac{d-1}{2}}\Gamma(\frac{2-d}{2d-2})
{\Gamma(\frac{d}{2d-2})}^{d-2}l^{d-1}L^{d-2}}{(2d-2){\Gamma(\frac{1}{2d-2})}^{d-1}y_a^{d-2}}
\bigg].
\eal
Combining (\ref{RTa}) with (\ref{Sdef}), we rewrite the generalized entropy in terms of $w_a$ as
\begin{equation}
\begin{split}
S_{\text{gen}}=\frac{1}{4G_N^{(d+1)}}\bigg\{ &\frac{l^{d-1}}{d-2}\left(\frac{L}{\epsilon}\right)^{d-2}+\frac{\sqrt{\pi}}{2d-2}\frac{\Gamma(\frac{2-d}{2d-2})}{\Gamma(\frac{1}{2d-2})}l^{d-1}(Lv_a)^{d-2}\\
&+\eta\frac{l^{d-1}L^{d-2}v_a^{d-2}}{d-2}\bigg[w_a^{2-d}F\left(\frac{1}{2},-\frac{d-2}{2(d-1)};\frac{d}{2(d-1)};w_a^{2d-2}\right)\\
&-F\left(\frac{1}{2},-\frac{d-2}{2(d-1)};\frac{d}{2(d-1)};1\right) \bigg]
+\frac{l^{d-1}}{d-2}\left(\frac{Ll v_a}{w_a
\epsilon_y}\right)^{d-2}\\
&+\frac{\pi^{\frac{d-1}{2}}\Gamma(\frac{2-d}{2d-2})
{\Gamma(\frac{d}{2d-2})}^{d-2}l^{d-1}L^{d-2}v_a^{d-2}{\cos\theta}^{d-2}}{(2d-2){\Gamma(\frac{1}{2d-2})}^{d-1}w_a^{d-2}}\ \bigg\}\ ,
\end{split}
\end{equation}
with $w_a$ to be varied. Then $S_{\text{DES}}$ is calculated as
\bal
S_{\text{DES}}=\min_{w_a}S_{\text{gen}}(w_a)\ .
\eal
Notice that unlike 2 dimensional BCFT, the entanglement due to the defect theory on the brane will shift the Ryu-Takayanagi surface.
\subsection{Boundary QES}\label{}
Now we compute the entropy of the strip from the boundary point of view, where the brane CFT is coupled to gravity. In this set-up, the gravitational region is at $x<0$ where we identify $x$ with $-y$, and the non-gravitational region is at $x>0$. So the metric is $ds^2=\Omega^{-2}(x)(d\tau^2+dx^2+d\vec{u}^2)$ in which the warped factor is
\begin{equation}
\begin{split}
\Omega (x)=\begin{cases}1,\quad &x>0\\
-\frac{x\cos \theta}{l},\quad &x<0
\end{cases}\ .
\end{split}
\end{equation}

 In the formula of the entropy, we will holographically calculate the matter term in an AdS$_\text{d+1}$ with the cut off at $z=\epsilon \Omega(x)$ when $x>0$ and $z=\epsilon_y \Omega(x)$ when $x<0$. The matter term receives a strip contribution on the brane with the boundaries at $(x_a,\tau_a)$ ($x_a<0$). This boundary brings an area term, i.e.
\begin{equation}
\label{atstr}
\begin{split}
S_{\text{area}}=& \frac{L^{d-2}}{4G_N^{(d)}}\left(\frac{l}{\cos \theta}\right)^{d-2}\frac{1}{|x_a|^{d-2}}\\
=&\frac{l^{d-1}}{4G_N^{(d+1)}}\frac{L^{d-2}}{|x_a|^{d-2}}\int_0^{\theta}\cos^{1-d}\theta' d\theta'\ .
\end{split}
\end{equation}
Note that in the second step the Newton constant on the brane $G_N^{(d)}$ has been replaced by the bulk one $G_N^{(d+1)}$ through partial Randall-Sundrum~\cite{Deng:2020ent}.

For the matter term $S_{\text{matter}}(x_a,\tau_a)$, holographically it can be calculated with an RT surface connecting $(x_1,\tau_1)$ with $(x_a,\tau_a)$. The time reflection symmetry guarantees $\tau_a=\tau_1$.

Now we write explicitly $S_{\text{matter}}(x_a,\tau_1)$. It is
\begin{equation}
\begin{split}
S_{\text{matter}}(x_a,\tau_1)=&\frac{1}{4G_N^{(d+1)}}\bigg[\frac{l^{d-1}}{d-2}L^{d-2}\left(\frac{1}{(\epsilon \Omega(x_1))^{d-2}}+\frac{1}{(\epsilon_y\Omega(x_a))^{d-2}}\right)\\
&-\frac{2^{d-1}\pi^{\frac{d-1}{2}}l^{d-1}}{d-2}\left(\frac{\Gamma (\frac{d}{2d-2})}{\Gamma (\frac{1}{2d-2})}\right)^{d-1}\left(\frac{L}{x_1-x_a}\right)^{d-2}\bigg]\\
=&\frac{1}{4G_N^{(d+1)}}\bigg[\frac{l^{d-1}}{d-2}L^{d-2}\left(\frac{1}{\epsilon^{d-2}}+\frac{l^{d-2}}{(\epsilon_y|x_a|\cos \theta)^{d-2}}\right)\\
&-\frac{2^{d-1}\pi^{\frac{d-1}{2}}l^{d-1}}{d-2}\left(\frac{\Gamma (\frac{d}{2d-2})}{\Gamma (\frac{1}{2d-2})}\right)^{d-1}\left(\frac{L}{x_1+|x_a|}\right)^{d-2}\bigg]\ .
\end{split}
\end{equation}
Then, the generalized entropy becomes
\begin{equation}
\label{genQESstr}
\begin{split}
S_{\text{gen}}(|x_a|)=&S_{\text{area}}+S_{\text{matter}}\\
=&\frac{l^{d-1}}{4G_R^{(d+1)}}\frac{L^{d-2}}{|x_a|^{d-2}}\int_0^{\theta}\cos^{1-d}\theta' d\theta'+\frac{l^{d-1}L^{d-2}}{4G_N^{(d+1)}}\bigg[\frac{1}{(d-2)\epsilon^{d-2}}\\
&-\frac{2^{d-1}\pi^{\frac{d-1}{2}}}{d-2}\left(\frac{\Gamma (\frac{d}{2d-2})}{\Gamma (\frac{1}{2d-2})}\right)^{d-1}\frac{1}{(x_1+|x_a|)^{d-2}}\bigg]\ ,
\end{split}
\end{equation}
where we have renormalized the Newton constant $G_R^{(d+1)}$ in the area term to incorporate the UV cutoff in the gravitational region. The extremization condition gives
\begin{equation}
\begin{split}
0=\partial_{|x_a|}S_{\text{gen}}(|x_a|)=&-\frac{l^{d-1}L^{d-2}}{4G_R^{(d+1)}}|x_a|^{1-d}(d-2)\int_0^{\theta}\cos^{1-d}\theta' d\theta'\\
&+\frac{l^{d-1}L^{d-2}}{4G_N^{(d+1)}}(x_1+|x_a|)^{1-d}2^{d-1}\pi^{\frac{d-1}{2}}\left(\frac{\Gamma (\frac{d}{2d-2})}{\Gamma (\frac{1}{2d-2})}\right)^{d-1}\ .
\end{split}
\end{equation}
And the solutions are
\begin{equation}
\begin{split}
\frac{1}{|x_{a,1}|}=0,\ \frac{1}{|x_{a,2}|}=\frac{1}{x_1}\left(\left(\frac{G_R^{(d+1)}}{G_N^{(d+1)}}\right)^{\frac{1}{d-1}}\frac{2\pi^{1/2}\frac{\Gamma (\frac{d}{2d-2})}{\Gamma (\frac{1}{2d-2})}}{\left((d-2)\int_0^{\theta}\cos^{1-d}\theta' d\theta'\right)^{\frac{1}{d-1}}}-1\right)\ .
\end{split}
\end{equation}
When the second solution $x_a=x_{a,2}$ exists, namely
\begin{equation}
\label{coan}
\begin{split}
(d-2)\int_0^{\theta}\cos^{1-d}\theta' d\theta'<\frac{G_R^{(d+1)}}{G_N^{(d+1)}}\left(2\pi^{1/2}\frac{\Gamma (\frac{d}{2d-2})}{\Gamma (\frac{1}{2d-2})}\right)^{d-1}\ ,
\end{split}
\end{equation}
we find $\partial_{|x_a|}S_{\text{gen}}(|x_a|)>0$ for $|x_a|\in (|x_{a,2}|,\infty)$. It implies the second solution is the minimal point. Plugging the solutions into (\ref{genQESstr}), we finally get
\begin{equation}
\begin{split}
S_{\text{QES}}=\begin{cases}\frac{l^{d-1}L^{d-2}}{4G_N^{(d+1)}}\frac{1}{(d-2)\epsilon^{d-2}},\quad &\theta>\theta_c\\
S_{\text{gen}}(|x_{a,2}|),\quad &\theta<\theta_c ,
\end{cases}
\end{split}
\end{equation}
with $\theta_c$ the critical angle of the inequality (\ref{coan}).
\subsection{Comparison between QES and DES}
\begin{figure}
  \centering
  \includegraphics[width=10cm]{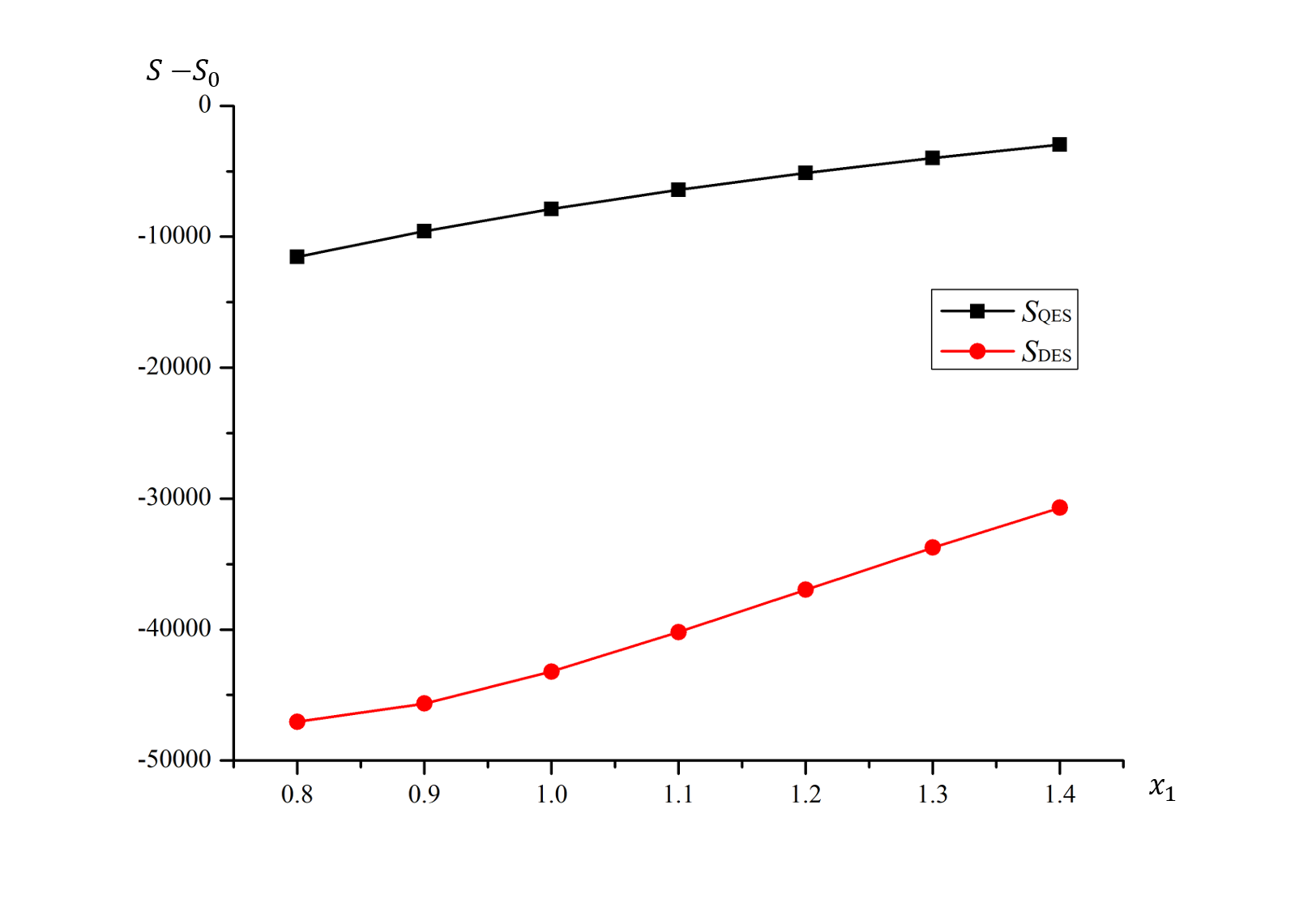}\\
  \caption{Entropy curve with respect to $x_1$, with $S_0=\frac{1}{4G_N^{(d+1)}} \frac{l^{d-1}}{d-2}\left(\frac{L}{\epsilon}\right)^{d-2}$ .}\label{15}
\end{figure}

\begin{figure}
  \centering
  \includegraphics[width=10cm]{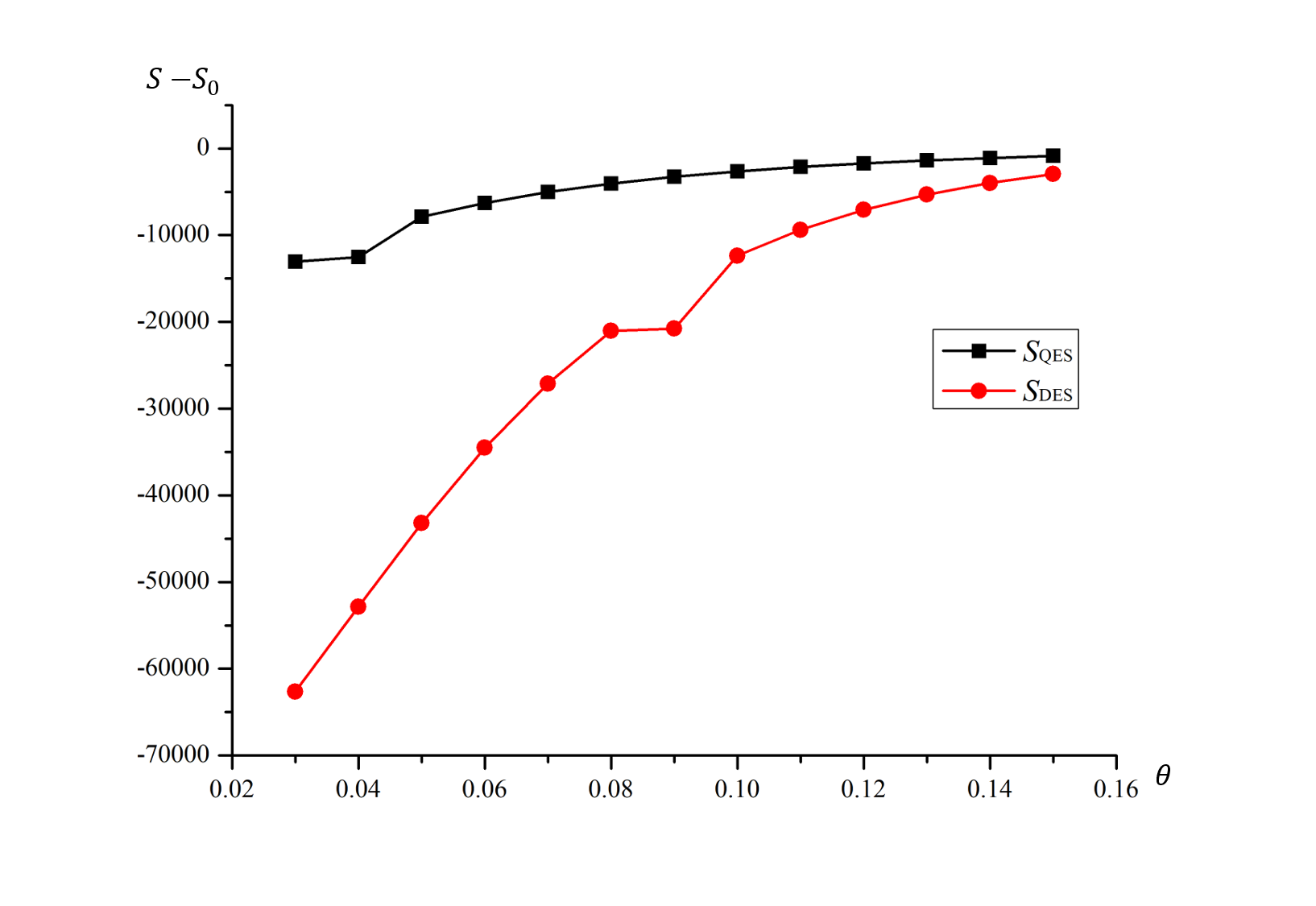}\\
  \caption{Entropy curve with respect to $x_1$, with $S_0=\frac{1}{4G_N^{(d+1)}} \frac{l^{d-1}}{d-2}\left(\frac{L}{\epsilon}\right)^{d-2}$ .}\label{22}
\end{figure}

In this section we compare the result obtained from DES and QES calculation. In QES side, one can get an analytical solution for the extremal point as well as the entropy. However, in general it is hard to get an analytical solution in DES side. Therefore we give numerical comparison for DES and QES. We choose $d=4$, $L=1000$, $l=1$, $G_N^{(5)}=G_R^{(5)}=1$, $\epsilon=10^{-3}$ and the variables are $x_1$ and $\theta$ respectively. Notice that comparing with boundary QES result, there is an extra free parameter $\epsilon_y$ in DES. Demanding that the island boundary in QES is the same as that in DES on the brane, one can fix the value of $\epsilon_y$. This matching condition is physical, because otherwise there will be a mismatch from the viewpoint of entanglement wedge. Thus by imposing the matching condition and choose $x_1$ and $\theta$ properly we get the numerical result of DES. The comparison between DES and QES when $\theta=0.05$ and $x_1\in[0.8,1.4]$ are shown in Fig.\ref{15}.
With $x_1=1$ fixed and $\theta\in[0.03,0.15]$ as the variable, the data and diagram for DES and QES are shown in Fig.\ref{22}.

From the numerical result, one can see that the entropy calculated by DES is always smaller than the entropy obtained from QES.
\section{Page curve for eternal black hole in BCFT$_{d}$}\label{sec5}
There is a higher-dimensional generalization of the eternal black hole in Section \ref{sec4}, where the BCFT$_{d}$ has a cylinder boundary $S\times R^{d-2}$. More specifically, the boundary is at $x^2_1+\tau^2=1$ with no restriction on $x_i,i=2,\cdots, d-1$. The AdS$_{d+1}$ dual has an EOW brane located at
\begin{equation}
\label{beq}
\begin{split}
(z+\tan\theta)^2+x^2_1+\tau^2=\sec^2\theta\ .
\end{split}
\end{equation}
This is the same equation as that in Section \ref{sec4}, but now the EOW brane has more dimensions $x_i,i=2,\cdots, d-1$. It is supported by certain stress energy tensor following the Neumann boundary condition (\ref{NBC}). Once the embedding function of the EOW brane is given, one can directly compute the induced metric and the extrinsic curvature, thus derive the stress energy tensor. One can also check null energy condition for the stress tensor. To illustrate, we give an example of $d=3$ with AdS radius taken to be $1$. The embedding function is
\begin{equation}
\begin{split}
f=(z+\tan\theta)^2+x^2_1-t^2-\sec^2\theta=0 .
\end{split}
\end{equation}
Thus the normal vector $n_{a}=\frac{\partial f}{\partial x^{a}}$ (toward the outside direction) is
\bal
n_{t}&=\frac{t\cos\theta}{z},\\
n_{z}&=-\frac{z\cos\theta+\sin\theta}{z},\\
n_{x_1}&=-\frac{x_1\cos\theta}{z},\\
n_{x_2}&=0.
\eal
The induced metric given by $h_{ab}=g_{ab}-n_{a}n_{b}$ is
\bal
h_{t t}&=\frac{1}{z^2}-\frac{t^2\cos^2\theta}{z^2},\\
h_{t x_1}=h_{x_1 t}&=\frac{tx_1\cos^2\theta}{z^2},\\
h_{t z}=h_{z t}&=\frac{t\cos\theta(z\cos\theta+\sin\theta)}{z^2},\\
h_{x_1 x_1}&=\frac{1}{z^2}-\frac{x_1^2\cos^2\theta}{z^2},\\
h_{x_1 z}=h_{zx_1}&=-\frac{x_1\cos\theta(z\cos\theta+\sin\theta)}{z^2},\\
h_{zz}&=\frac{1}{z^2}-\frac{(z\cos\theta+\sin\theta)^2}{z^2},\\
h_{x_2 x_2}&=\frac{1}{z^2},
\eal
with other components vanishing.
The extrinsic curvature can be calculated as $K_{ab}=h_{a}^{c} h_{b}^{d} \nabla_{c} n_{d}$ and $K=h^{ab} K_{ab}$. Then one can obtain the stress energy tensor as $T_{ab}=\frac{1}{8\pi G_{N}}\left[K_{a b}^{(h)}-h_{a b} K^{(h)}\right]$. Now we check whether this $T_{ab}$ satisfies null energy condition $T_{ab}N^a N^b\geq 0$, where $N^a$ is arbitrary null vector. In present case, one can choose $N^a=(1,0,\frac{t}{x},-\frac{\sqrt{x^2-t^2}}{x})$. It's easy to check that $N^a$ satisfies $N^an_a=0$ and $N^aN_a=0$, thus it is indeed a null vector on the brane. One can therefore check~\footnote{The stress tensor $T_{ab}$ is computed by mathematica code which has partially used the package named {\it diffgeo.m} written by  Matthew Headrick, see \cite{Matthew Headrick}.}
\bal\begin{split}
T_{ab}N^a N^b=\frac{(x-t) (t+x)\cos\theta}{x^2 z }\geq 0.
\end{split}\eal
By noting that the null vector is real, this inequality is true for $\text{$\theta$}\in[0,\frac{\pi}{2}]$ and $z>0$.

The cut off where the BCFT lives is $z=\epsilon$. Like what we did in Section \ref{sec4}, now we consider a subregion of the bath bounded by $(\tau,x_1)=(\tau_0,x_0)$ and $(\tau_0,-x_0)$, with the other coordinates freely extended. To proceed the calculation of the entanglement entropy, we will cut off such that $-L/2<x_i<L/2, i=2,\cdots ,d-1$. Similar to Section \ref{sec4}, eventually we will rotate to coordinate system $(T,R)$ with the transformations given by
\begin{equation}
\label{ctr}
\begin{split}
x_1=e^R\cosh T,\ \tau=ie^R\sinh T\ .
\end{split}
\end{equation}

\subsection{Early-time phase}
Now we compute the entropy for the chosen subregion of the bath. There are two possible phases, i.e., the connected phase and the disconnected phase. The connected one does not include contribution from the brane, so the entropy is just the matter entropy, i.e.~\cite{Ryu:2006ef}
\begin{equation}
\label{eph}
\begin{split}
S(R_0,T)&=\frac{1}{4G_N^{(d+1)}}\bigg[\frac{2l^{d-1}}{d-2}\frac{L^{d-2}}{\epsilon^{d-2}}-\frac{2^{d-1}\pi^{\frac{d-1}{2}}l^{d-1}}{d-2}\left(\frac{\Gamma (\frac{d}{2d-2})}{\Gamma (\frac{1}{2d-2})}\right)^{d-1}\left(\frac{L}{2x_0}\right)^{d-2}\bigg]\\
&=\frac{1}{4G_N^{(d+1)}}\bigg[\frac{2l^{d-1}}{d-2}\frac{L^{d-2}}{\epsilon^{d-2}}-\frac{2^{d-1}\pi^{\frac{d-1}{2}}l^{d-1}}{d-2}\left(\frac{\Gamma (\frac{d}{2d-2})}{\Gamma (\frac{1}{2d-2})}\right)^{d-1}\left(\frac{L}{2e^{R_0}\cosh T}\right)^{d-2}\bigg]\ .
\end{split}
\end{equation}
\subsection{Late-time phase}
In the disconnected phase, we only consider the entropy without defect contribution, i.e. the RT surface. The RT surface ends on the brane at $(\tau_a,-x_{1,a},z_a)$ and $(\tau_a,x_{1,a},z_a)$. Since the two disconnected RT surfaces are identical, we will just look at one of them, e.g. the one ending on $(\tau_a,x_{1,a},z_a)$ and $(\tau_0,x_0,\epsilon)$. With the coordinate transformations $\tau=r\sin T_E$ and $x_1=r\cos T_E$~\footnote{By comparing it with (\ref{ctr}), one can find the relation that $r=e^R$ and $T_E=iT$.}, we rewrite the endpoint on the brane as $(T_{E,a},r_a,z_a)$. Note that $z_a$ is a function of $r_a$ from the equation (\ref{beq}), more specifically,
\begin{equation}
\begin{split}
z_a=\sqrt{\sec^2\theta-r_a^2}-\tan\theta\ .
\end{split}
\end{equation}
From the symmetry of $T_E \to 2T_E(\tau_0,x_0)-T_E$, we deduce that the RT surface is in the slice $T_E=T_E(\tau_0,x_0)$. To be the minimal, the RT surface should intersect the brane perpendicularly, which gives the condition that
\begin{equation}
\label{ranc}
\begin{split}
\frac{dz}{dx}&=\frac{z_a+\tan\theta}{r_a}\\
&=\frac{\sqrt{\sec^2\theta-r_a^2}}{r_a}\ .
\end{split}
\end{equation}
Combining it with the first equation in (\ref{dzdx}), we solve that
\begin{equation}
\begin{split}
z_*=z_a\left(\frac{\sec \theta}{r_a}\right)^{\frac{1}{d-1}}\ .
\end{split}
\end{equation}
And similar to (\ref{RTa}), the area of the RT surface is
\begin{equation}
\label{dcph}
\begin{split}
A=&l^{d-1}L^{d-2}\bigg\{\frac{1}{(d-2)\epsilon^{d-2}}+\frac{\sqrt{\pi}}{2d-2}\frac{\Gamma(\frac{2-d}{2d-2})}{\Gamma(\frac{1}{2d-2})z_*^{d-2}}\\
&+\frac{1}{z_*^{d-2}(d-2)}\bigg[\left(\frac{z_a}{z_*}\right)^{2-d}F\left(\frac{1}{2},-\frac{d-2}{2(d-1)};\frac{d}{2(d-1)};\left(\frac{z_a}{z_*}\right)^{2d-2}\right)\\
&-F\left(\frac{1}{2},-\frac{d-2}{2(d-1)};\frac{d}{2(d-1)};1\right) \bigg]\bigg\}\ .
\end{split}
\end{equation}
The value of $r_a$ can be determined by solving an equation as in (\ref{dzdx}), namely
\begin{equation}
\begin{split}
r_1=&-\frac{z_*}{d}\bigg[ \left(\frac{z_a}{z_*}\right)^dF\left(\frac{1}{2},\frac{d}{2(d-1)};\frac{3d-2}{2(d-1)};\left(\frac{z_a}{z_*}\right)^{2(d-1)}\right) \\
&-F\left(\frac{1}{2},\frac{d}{2(d-1)};\frac{3d-2}{2(d-1)};1\right)  \bigg]+\frac{\sqrt{\pi}\Gamma(\frac{d}{2d-2})}{\Gamma(\frac{1}{2d-2})}z_*+r_a \ .
\end{split}
\end{equation}
Note that (\ref{dcph}) has no dependence on time $T$. Finally, the entropy in this phase is
\begin{equation}
\label{lph}
\begin{split}
S(R_0,T)=\frac{A}{2G_N^{(d+1)}} \ .
\end{split}
\end{equation}
	\begin{figure}[htbp]
	\centering
	\includegraphics[width=10cm]{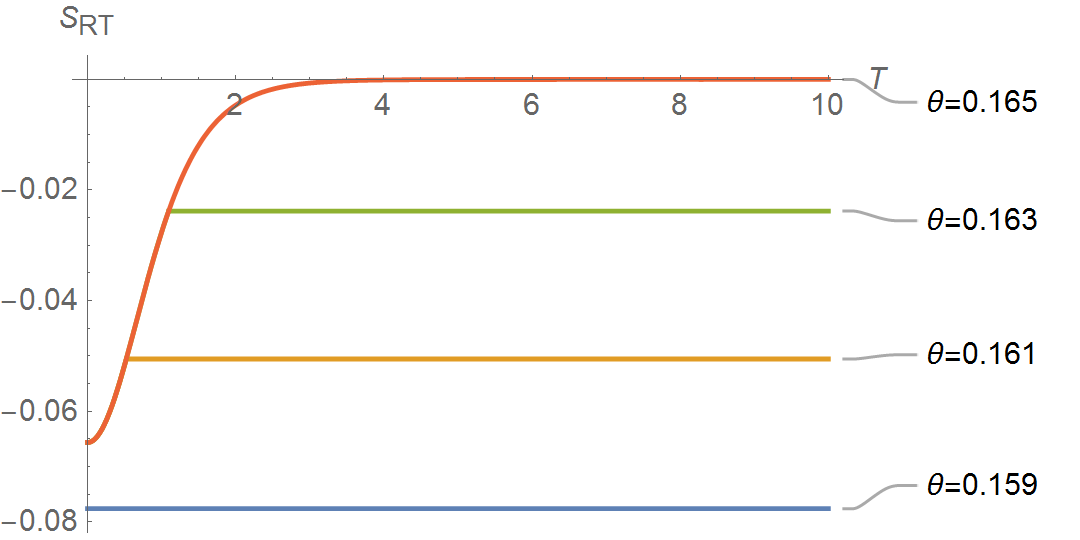}\\
	\caption{The entropy (in the unit of $\frac{l^{d-1}L^{d-2}}{4G_N^{(d+1)}}$) with respect to time $T$ for $d=4$, $X_0=0.1$ and $\theta=0.159,0.161,0.163,0.165$. We also substract the constant term $\frac{2l^{d-1}}{d-2}\frac{L^{d-2}}{\epsilon^{d-2}}$.}
	\label{hdpc}
	\end{figure}
	
In Fig.\ref{hdpc}, we plot the entropy which is the minimum of (\ref{eph}) and (\ref{lph}). We find that the phase transitions only occur in a small range of $\theta$. When $\theta$ is too small, e.g. $\theta=0.159$, the late-time phase (\ref{lph}) dominates in the beginning and the entropy remains constant. When $\theta$ is too large, e.g. $\theta=0.165$, the early-time phase (\ref{eph}) dominates in the whole time period.
\section{Entanglement entropy for a ball in BCFT$_{d}$}\label{sec3}
Now we consider a $d-1$-dimensional time slice of BCFT$_{d}$ which has a spherical boundary $r'=1$, where $\vec{r'}=(x'_1,x'_2,\cdots x'_{d-1})$. The holographic dual of the BCFT is an AdS$_{d+1}$ with an EOW brane located at $(z'+\tan\theta)^2+r'^2=\sec^2\theta$. The cut off where the BCFT lives is $z'=\epsilon$. Note that under the conformal transformations
\begin{equation}
\label{sctsph}
\begin{split}
x_1&=\frac{2(r'^2+z'^2-1)}{r'^2+z'^2+2x'_1+1}\\
x_{i>1}&=\frac{4x'_{i>1}}{r'^2+z'^2+2x'_1+1}\\
z&=\frac{4z'}{r'^2+z'^2+2x'_1+1}\ ,
\end{split}
\end{equation}
which preserves the metric, the boundary is mapped to a $d-2$-dimensional hyperplane $x_1=0$ and the EOW brane is mapped to a $d-1$-dimensional hyperplane $x_1=-z\tan \theta$. In the rest of this section, we will calculate the entanglement entropy of a subregion bounded by a $d-2$-sphere $r'=r'_0$ and we take $d=4$ as an example.

\subsection{Bulk DES}\label{}
\begin{figure}[h]
\centering
 \includegraphics[width=13cm,height=8cm]{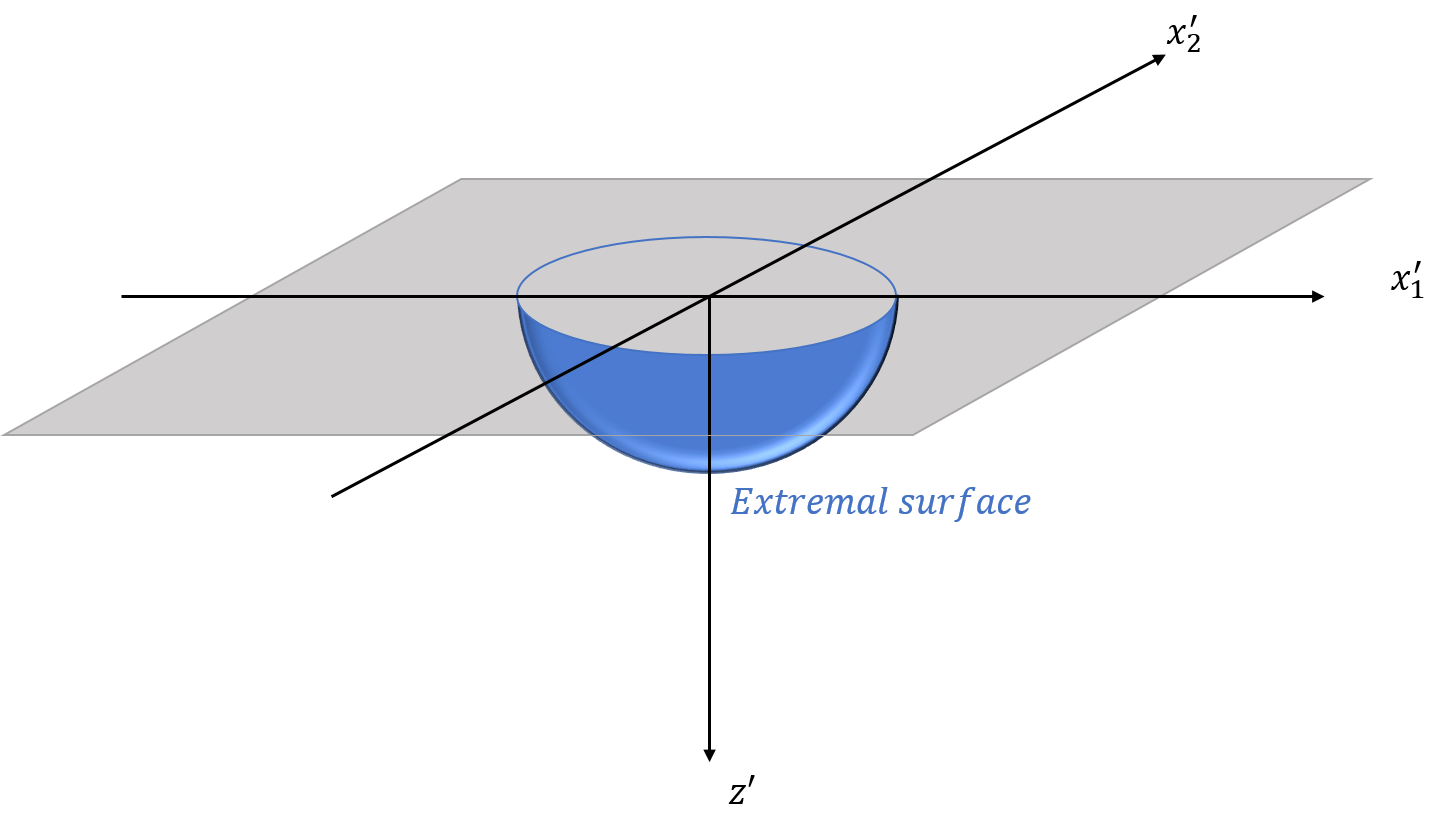}\\
 \caption{The phase that extremal surface does not intersect with the EOW brane.}\label{8}
\end{figure}

The proposal of defect extremal surface formula is (\ref{DES}). In general there are two phases of the extremal surface. One does not intersect with the EOW brane while the other does. In the former phase, no defect contribution would be included as shown in Fig.\ref{8}, and the entropy is given by the RT surface~\cite{Ryu:2006ef}
\begin{equation}
\label{cDESsph}
\begin{split}
S_{\text{DES}}=\frac{\pi l^3}{2G_N^{(5)}}\left(\left(\frac{r'_0}{\epsilon}\right)^2-\log \frac{r'_0}{\epsilon}\right)\ .
\end{split}
\end{equation}
For the phase where the extremal surface intersects with the brane, one should also add the contribution of CFT matter on the brane to the generalized entropy functional. We assume that the extremal surface intersects the brane at $(z'_1,r'_1)$ where $z'_1=\sqrt{\sec^2 \theta-r'^2_1}-\tan \theta$ according to the equation for the EOW brane. Then the area of a tube anchoring on $r'= r'_0$ and $r'=r'_1$ is given by the functional
\begin{equation}
\label{RTsph0}
\begin{split}
A=4\pi l^3\int_{r'_1}^{r'_0}dr'\sqrt{1+\left(\frac{dz'}{dr'}\right)^2}\frac{r'^2}{z'^3}\ ,
\end{split}
\end{equation}
where $z'$ is a function of $r'$.
With the change of variables $r'=\frac{e^{\zeta}P}{\sqrt{1+P^2}}$ and $z'=\frac{e^{\zeta}}{\sqrt{1+P^2}}$~\footnote{Note that the following analysis of the tube RT surface is similar to~\cite{Krtous:2014pva}, where $d=2$ though.}, the endpoints $(r_0',\epsilon)$ and $(r_1',z_1')$ are mapped to
\begin{equation}
\begin{split}
P_0=\frac{r'_0}{\epsilon}\to \infty ,& \quad \zeta_0=\log r_0'\ ,\\
P_1=\frac{r'_1}{z'_1},& \quad \zeta_1=\frac{1}{2}\log (r'^2_1+z'^2_1)\ ,
\end{split}
\end{equation}
and the functional becomes
\begin{equation}
\label{RTsph}
\begin{split}
A=4\pi l^3\left(\int_{P_*}^{P_0}+\eta \int_{P_*}^{P_1}\right)dP\sqrt{1+(P^2+1)^2\left(\frac{d\zeta}{dP}\right)^2}\frac{P^2}{\sqrt{1+P^2}}\ ,
\end{split}
\end{equation}
where $P_{*}$ is the turning point of the RT surface or its extension and $\eta=\pm 1$ depending on whether $\zeta (P_*)>\zeta_1$ or not.

Since the Lagrangian has no explicit dependence on $\zeta$, it gives a constant of motion
\begin{equation}
\label{dzdP}
\begin{split}
P_{*}^2\sqrt{1+P^2_{*}}=\frac{P^2(1+P^2)^{3/2}\frac{d\zeta}{dP}}{\sqrt{1+(1+P^2)^2\left(\frac{d\zeta}{dP}\right)^2}}\ .
\end{split}
\end{equation}
Therefore,
\begin{equation}
\label{zeta'PsphRT}
\begin{split}
\zeta_0-\zeta_1=\left(\int_{P_*}^{\infty}+\eta\int_{P_*}^{P_1} \right) dP\frac{P_*^2\sqrt{1+P_*^2}}{(1+P^2)\sqrt{P^4(1+P^2)-P_*^4(1+P_*^2)}}\ ,
\end{split}
\end{equation}
from which $P_*$ can be solved with given $r'_1$ (Note that in most cases there are more than one solutions, and we should pick the one which gives the smallest RT surface). Furthermore, from this formula we can determine $\eta$, i.e.
\begin{equation}
\begin{split}
\eta=\begin{cases}1,\quad &\int_{P_*}^{\infty} dP\frac{P_*^2\sqrt{1+P_*^2}}{(1+P^2)\sqrt{P^4(1+P^2)-P_*^4(1+P_*^2)}}<\zeta_0-\zeta_1\\
-1, \quad & \text{otherwise}.
\end{cases}
\end{split}
\end{equation}
By substituting (\ref{dzdP}) back into (\ref{RTsph}) we can calculate the RT term
\begin{equation}
\label{RTsph2}
\begin{split}
S_{\text{RT}}(r_1')&=\frac{A}{4G_N^{(5)}}\\
&=\frac{\pi l^3}{G_N^{(5)}}\left(\int_{P_*}^{P_0}+\eta\int_{P_*}^{P_1}\right)dP\frac{P^4}{\sqrt{P^4(1+P^2)-P_*^4(1+P_*^2)}}\ .
\end{split}
\end{equation}

For the defect entropy, namely the entropy of the brane subregion bounded by $(z'_1,r'_1)$, we compute it holographically. The curved background can be recovered by picking the geodesic cut-off properly. To see this, we first look at the induced metric $ds^2_Q$ on the brane. Similar to the second term in (\ref{2dmet}), with the coordinate transformation
\begin{equation}
\label{rQ}
\begin{split}
x_1&=y\sin \theta\\
-z&=y\cos \theta\ ,
\end{split}
\end{equation}
we can write the induced metric as
\begin{equation}
\label{mQ}
\begin{split}
ds^2_Q=\frac{l^2}{\cos^2 \theta}\frac{dy^2+dx_2^2+dx_3^2}{y^2}\ .
\end{split}
\end{equation}
Now, to recover the metric (\ref{mQ}) from a dual AdS$_4$, we pick the cut-off at
\begin{equation}
\label{zQcof}
\begin{split}
z_Q=-\epsilon_y y\cos \theta\ ,
\end{split}
\end{equation}
where $z_Q$ denotes the radial coordinate of the dual AdS$_4$.

To calculate the defect entropy, it is convenient to do in prime coordinates since the boundary of the brane subregion is spherical. Similar to (\ref{sctsph}), we use the coordinate transformations
\begin{equation}
\label{sctQ}
\begin{split}
y&=\frac{2(r'^2_Q+z'^2_Q-1)}{r'^2_Q+z'^2_Q+2x'_{1,Q}+1}\\
x_{i>1}&=\frac{4x'_{i>1,Q}}{r'^2_Q+z'^2_Q+2x'_{1,Q}+1}\\
z_Q&=\frac{4z'_Q}{r'^2_Q+z'^2_Q+2x'_{1,Q}+1}\ ,
\end{split}
\end{equation}
in which $\vec{r'}_Q=(x_{1,Q}',x_{2,Q}',x_{3,Q}')$. With the coordinate changes, the cut-off (\ref{zQcof}) becomes~\footnote{Or
$P_{\epsilon_y}(r')=\frac{2r'_Q}{\epsilon_y\cos \theta (1-r'^2_Q)}$
in coordinates $(P_Q,\zeta_Q)$ with $r'_Q=\frac{e^{\zeta_Q}P_Q}{\sqrt{1+P^2_Q}}$ and $z'_Q=\frac{e^{\zeta_Q}}{\sqrt{1+P^2_Q}}$ (we will ignore the label $Q$ for $(P_Q,\zeta_Q)$ below).}
\begin{equation}\label{gco}
\begin{split}
z'_Q=\frac{\epsilon_y \cos \theta (1-r'^2_Q) }{2}\ .
\end{split}
\end{equation}
Note that by combining (\ref{sctsph}), (\ref{rQ}) and (\ref{sctQ}), we can solve that on the EOW brane ($z'_Q\to 0$) the relation between $\vec{r'}$ and $\vec{r'}_Q$ is
\begin{equation}
\label{rdrq}
\begin{split}
\vec{r'}&=\frac{2\vec{r'}_Q}{1+r'^2_Q+(1-r'^2_Q)\sin \theta}\ .
\end{split}
\end{equation}

\begin{figure}[h]
  \centering
  \includegraphics[width=10cm]{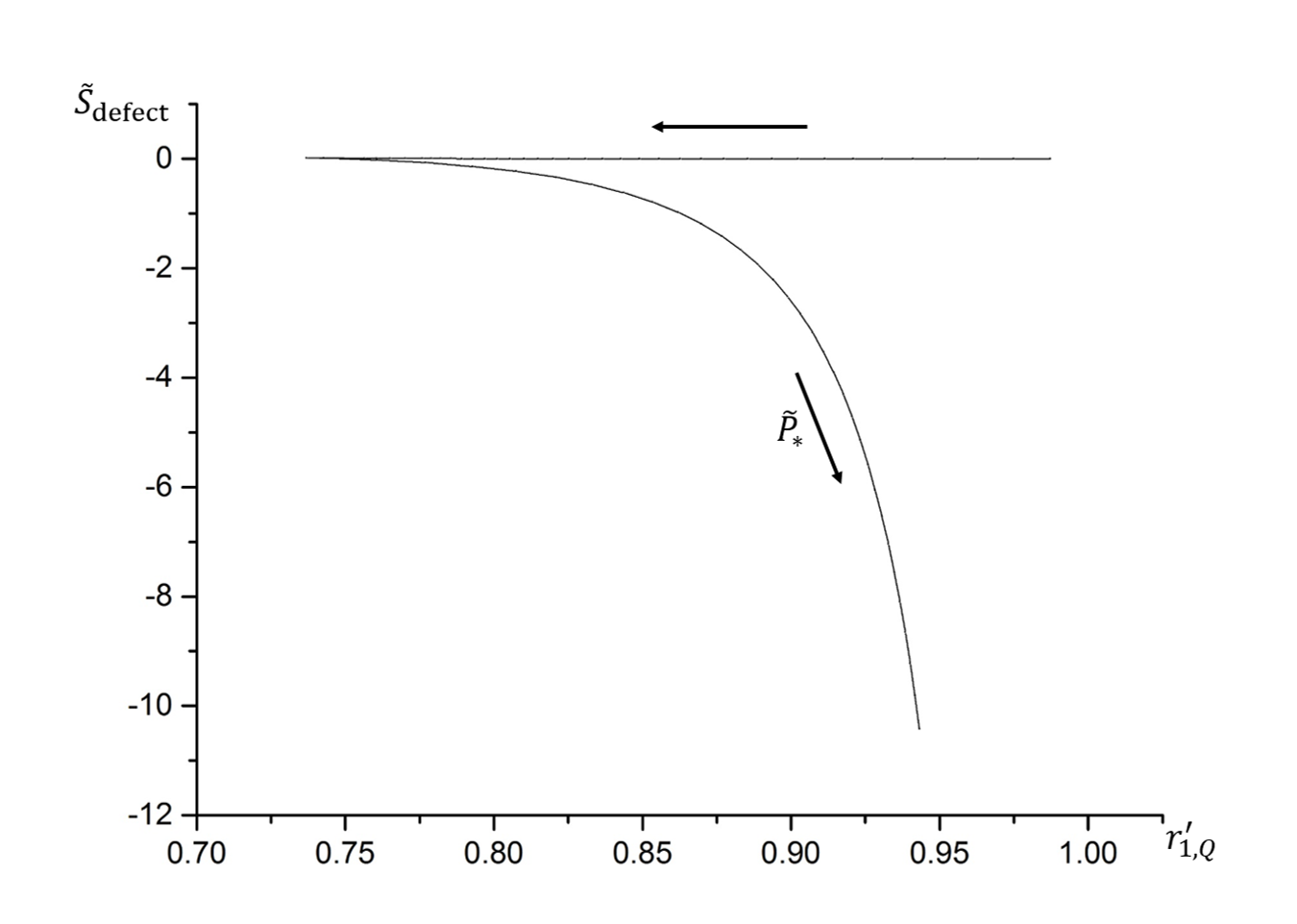}\\
  \caption{Plot of defect entropy (\ref{de1}) (subtracting the UV divergent terms which does not depend on $\tilde{P_*}$ explicitly) in the unit of $\frac{\pi l^3}{G_N^{(5)}}$ with respect to $r_{1,Q}'$. The arrows denote the increasing direction of $\tilde{P_*}$.}\label{Sde}
\end{figure}

The matter on the brane is a BCFT on curved background with zero boundary entropy, thus from AdS/BCFT one can determine the location of the bulk brane of this BCFT to be $z'^2_Q+r'^2_Q=1$ (i.e. $\theta=0$). When changed to $(P,\zeta)$ coordinate, it is simply $\zeta=0$. Like flat space BCFT, the RT surface have two phases, one doesn't intersect with the bulk brane, the other does. For the former phase, one can simply use the geodesic cut off (\ref{gco}) to replace the flat space cut off in (\ref{cDESsph}) and the result is
\bal
S_{\text{defect}}^{(0)}=\frac{\pi l^3}{2G_N^{(5)}}\left(\left(\frac{2r'_{1,Q}}{\epsilon_y\cos\theta(1-{r'_{1,Q}}^2)}\right)^2-\log \frac{2r'_{1,Q}}{\epsilon_y\cos\theta(1-{r'_{1,Q}}^2)}\right)\ .
\eal
For the later phase, by noticing that RT surface is orthogonal to the bulk brane at their intersection point, the derivative $\frac{dP}{d\zeta}$ at the intersection point is determined to be zero which means that the endpoint of RT surface is just its turning point. Thus, similar to (\ref{RTsph2})
\begin{equation}
\label{de1}
\begin{split}
S_{\text{defect}}=&\frac{\tilde A}{4G_N^{(5)}}\\
=&\frac{\pi l^3}{G_N^{(5)}}\int_{\tilde{P_*}}^{P_{\epsilon_y}(r'_1)}dP\frac{P^4}{\sqrt{P^4(1+P^2)-\tilde{P_*}^4(1+\tilde{P_*}^2)}}\\
=&\frac{\pi l^3}{G_N^{(5)}}\bigg[\int_{\tilde{P_*}}^{\infty}dP\left(\frac{P^4}{\sqrt{P^4(1+P^2)-\tilde{P_*}^4(1+\tilde{P_*}^2)}}-\frac{P^2}{\sqrt{1+P^2}}\right)\\
&+\int_{\tilde{P_*}}^{P_{\epsilon_y}(r'_1)}dP\frac{P^2}{\sqrt{1+P^2}}\bigg]\\
=&\frac{\pi l^3}{G_N^{(5)}}\bigg[\int_{\tilde{P_*}}^{\infty}dP\left(\frac{P^4}{\sqrt{P^4(1+P^2)-\tilde{P_*}^4(1+\tilde{P_*}^2)}}-\frac{P^2}{\sqrt{1+P^2}}\right)\\
&-\frac{1}{2}\tilde{P_*}\sqrt{1+\tilde{P_*}^2}+\frac{1}{2}\text{arcsinh}\tilde{P_*}\\
&+\frac{1}{2}\left(\frac{2r'_{1,Q}}{\epsilon_y\cos\theta(1-{r'_{1,Q}}^2)}\right)^2-\frac{1}{2}\log \frac{2r'_{1,Q}}{\epsilon_y\cos\theta(1-{r'_{1,Q}}^2)}\bigg]\\
=&\tilde{S}_{\text{defect}}+S_{\text{defect}}^{(0)}\ ,
\end{split}
\end{equation}
where $\tilde{P_*}$ denotes the turning point of bulk surface that ends at $P_1$ and $\tilde{P_*}$. In the last step we have extracted the finite part $\tilde{S}_{\text{defect}}(r'_1)$. Notice that by integrating the constant of motion (\ref{dzdP}), we can determine $\tilde{P_*}$ from $r'_{1,Q}$
\begin{equation}\label{tP*}
\begin{split}
r'_{1,Q}=\exp \left(-\int_{\tilde{P_*}}^{\infty} dP\frac{\tilde{P_*}^2\sqrt{1+\tilde{P_*}^2}}{(1+P^2)\sqrt{P^4(1+P^2)-\tilde{P_*}^4(1+\tilde{P_*}^2)}}\right)\ ,
\end{split}
\end{equation}
which in general is not single-valued. However, as shown in Fig.\ref{Sde}, the larger value of $\tilde{P_*}$ gives a smaller defect entropy. Also note that when $r_{1,Q}'<0.73693$ this phase disappears.

Then the defect entropy is given by
\bal
S_{\text{defect}}=S_{\text{defect}}^{(0)}+\min \left\{0,\tilde{S}_{\text{defect}}\right\}\ .
\eal
And the generalized entropy is
\bal\label{gensphDES}\begin{split}
S_{\text{gen}}(r_1')&=S_{\text{RT}}(r_1')+S_{\text{defect}}\ .
\end{split}\eal
By extremizing the generalized entropy functional, the entropy is eventually obtained as
\begin{equation}
\label{DESsph}
\begin{split}
S_{\text{DES}}=\min \left\{\frac{\pi l^3}{2G_N^{(5)}}\left(\left(\frac{r'_0}{\epsilon}\right)^2-\log \frac{r'_0}{\epsilon}\right) ,\min_{r_1'} \{ S_{\text{gen}}(r_1')\} \right\}\ .
\end{split}
\end{equation}

\subsection{Boundary QES}\label{}
Now we rederive the entropy of the same subregion from the boundary point of view. In the boundary description, the brane CFT matter complements the BCFT in the flat region with a transparent boundary condition. More specifically, we redefine that $x_1=y$ when $x_1<0$.

We compute the entropy of CFT holographically. By tuning the cut-off of the dual AdS, we can recover the CFT on the curved background. Just the same as (\ref{gco}), in the coordinate system $(r',z')$ the cut-off for the gravitational region $r'<1$ is
\begin{equation}
\begin{split}
z'=\frac{\epsilon_y \cos \theta (1-r'^2) }{2}\ .
\end{split}
\end{equation}

Similar to DES, there are two possible phases in the QES computation, one of which contains no contribution from the brane while the other does. Without contribution from the brane, the entropy is just the matter entropy~\cite{Ryu:2006ef}
\bal\label{nispsph}
S_{\text{QES}}&=S_{\text{matter}}(r'_0)\\
&=\frac{\pi l^3}{2G_N^{(5)}}\left(\left(\frac{r'_0}{\epsilon}\right)^2-\log \frac{r'_0}{\epsilon}\right)\ ,
\eal
which is the same as (\ref{cDESsph}).

Since the brane CFT is coupled to gravity, there is also a possibility that the matter term receives a subregion contribution on the brane bounded by a $2$-sphere $r'=r'_1$. The boundary will also bring an area term, i.e.
\bal
\label{atsph}
S_{\text{area}}&=\frac{\pi r'^2_1}{G_N^{(4)}}\left(\frac{2l}{(1-r'^2_1)\cos\theta}\right)^2\\
&=\frac{2\pi r'^2_1}{G_N^{(5)}}\frac{l^3}{(1-r'^2_1)^2}\left(\text{arctanh}\sin\theta+\frac{\sin \theta}{\cos^2\theta}\right)\ .
\eal
Note that in the second step the Newton constant on the brane $G_N^{(4)}$ has been replaced by the bulk one $G_N^{(5)}$ through the partial Randall-Sundrum~\cite{Deng:2020ent}.

For the matter term $S_{\text{matter}}(r'_1)$, holographically there are two phases of the RT surface. It is either disconnected or connected. The former phase should be abandoned because it is strictly larger than (\ref{nispsph}).
For the later case, the area of a tube anchoring on $r'= r'_0$ and $r'=r'_1$ is given by the functional
\begin{equation}
\begin{split}
A=4\pi l^3\int_{r'_1}^{r'_0}dr'\sqrt{1+\left(\frac{dz'}{dr'}\right)^2}\frac{r'^2}{z'^3}\ ,
\end{split}
\end{equation}
which is the same as (\ref{RTsph0}) in the bulk calculation (although the boundary $z'(r'_1)$ is different). Similarly, we minimize the functional and get the matter term
\begin{equation}
\begin{split}
S_{\text{matter}}(P_*)&=\frac{A}{4G_N^{(5)}}\\
=&\frac{\pi l^3}{G_N^{(5)}}\left(\int_{P_*}^{P_0}+\int_{P_*}^{P_{\epsilon_y}}\right)dP\frac{P^4}{\sqrt{P^4(1+P^2)-P_*^4(1+P_*^2)}}\\
=&\frac{2\pi l^3}{G_N^{(5)}}\int_{P_*}^{\infty}dP\left(\frac{P^4}{\sqrt{P^4(1+P^2)-P_*^4(1+P_*^2)}}-\frac{P^2}{\sqrt{1+P^2}}\right)\\
&+\frac{\pi l^3}{G_N^{(5)}}\left(\int_{P_*}^{P_0}+\int_{P_*}^{P_{\epsilon_y}}\right)dP \frac{P^2}{\sqrt{1+P^2}}\\
=&\frac{2\pi l^3}{G_N^{(5)}}\int_{P_*}^{\infty}dP\left(\frac{P^4}{\sqrt{P^4(1+P^2)-P_*^4(1+P_*^2)}}-\frac{P^2}{\sqrt{1+P^2}}\right)\\
&+\frac{\pi l^3}{2G_N^{(5)}}\left(\frac{r'^2_0}{\epsilon^2}-\log \frac{r'_0}{\epsilon}+\frac{4r'^2_1}{\epsilon_y^2\cos^2\theta (1-r'^2_1)^2}-\log \frac{2r'_1}{\epsilon_y\cos \theta (1-r'^2_1)}\right)\\
&-\frac{\pi l^3}{G_N^{(5)}}(P_*\sqrt{1+P^2_*}-\text{arcsinh} P_* )\ .
\end{split}
\end{equation}
where $P_{0({\epsilon_y})}$ is the cut-off at $r'= r'_{0(1)}$, i.e.
\begin{equation}
\begin{split}
P_0&=\frac{r'_0}{\epsilon}\ ,\\
P_{\epsilon_y}&=\frac{2r'_1}{\epsilon_y\cos \theta (1-r'^2_1)}\ ,
\end{split}
\end{equation}
and $r'_1$ can be expressed in terms of $P_*$ by integrating (\ref{dzdP})
\begin{equation}
\begin{split}
r'_1=r'_0\exp \left(-2\int_{P_*}^{\infty} dP\frac{P_*^2\sqrt{1+P_*^2}}{(1+P^2)\sqrt{P^4(1+P^2)-P_*^4(1+P_*^2)}}\right)\ .
\end{split}
\end{equation}
Combined with the area term, it gives the generalized entropy
\begin{equation}
\begin{split}
 S_{\text{gen}}(P_*)=&S_{\text{area}}+S_{\text{matter}}\\
 =&\frac{2\pi r'^2_1}{G_R^{(5)}}\frac{l^3}{(1-r'^2_1)^2}\left(\text{arctanh}\sin\theta+\frac{\sin \theta}{\cos^2\theta}\right)\\
 &+\frac{2\pi l^3}{G_N^{(5)}}\int_{P_*}^{\infty}dP\left(\frac{P^4}{\sqrt{P^4(1+P^2)-P_*^4(1+P_*^2)}}-\frac{P^2}{\sqrt{1+P^2}}\right)\\
&-\frac{\pi l^3}{G_N^{(5)}}(P_*\sqrt{1+P^2_*}-\text{arcsinh} P_* )+\frac{\pi l^3}{2G_N^{(5)}}\left(\frac{r'^2_0}{\epsilon^2}-\log \frac{r'_0}{\epsilon}\right)\ .
\end{split}
\end{equation}
Note that we have renormalized the Newton constant $G_N^{(5)}$ to $G_R^{(5)}$. To get the final entropy, we need to extremize $S_{\text{gen}}(P_*)$ with respect to $P_*$. To summarize,
\begin{equation}
\begin{split}
S_{\text{QES}}=\frac{\pi l^3}{2G_N^{(5)}}\left(\frac{r'^2_0}{\epsilon^2}-\log \frac{r'_0}{\epsilon}\right)+\min \left\{0, \min_{P_*}\tilde{S}_{\text{gen}}(P_*)\right\}\ ,
\end{split}
\end{equation}
where $\tilde{S}_{\text{gen}}(P_*)$ denotes $S_{\text{gen}}(P_*)$ with the last term subtracted.
\subsection{Comparison between QES and DES}\label{}
\begin{figure}[h]
  \centering
  \includegraphics[width=9cm]{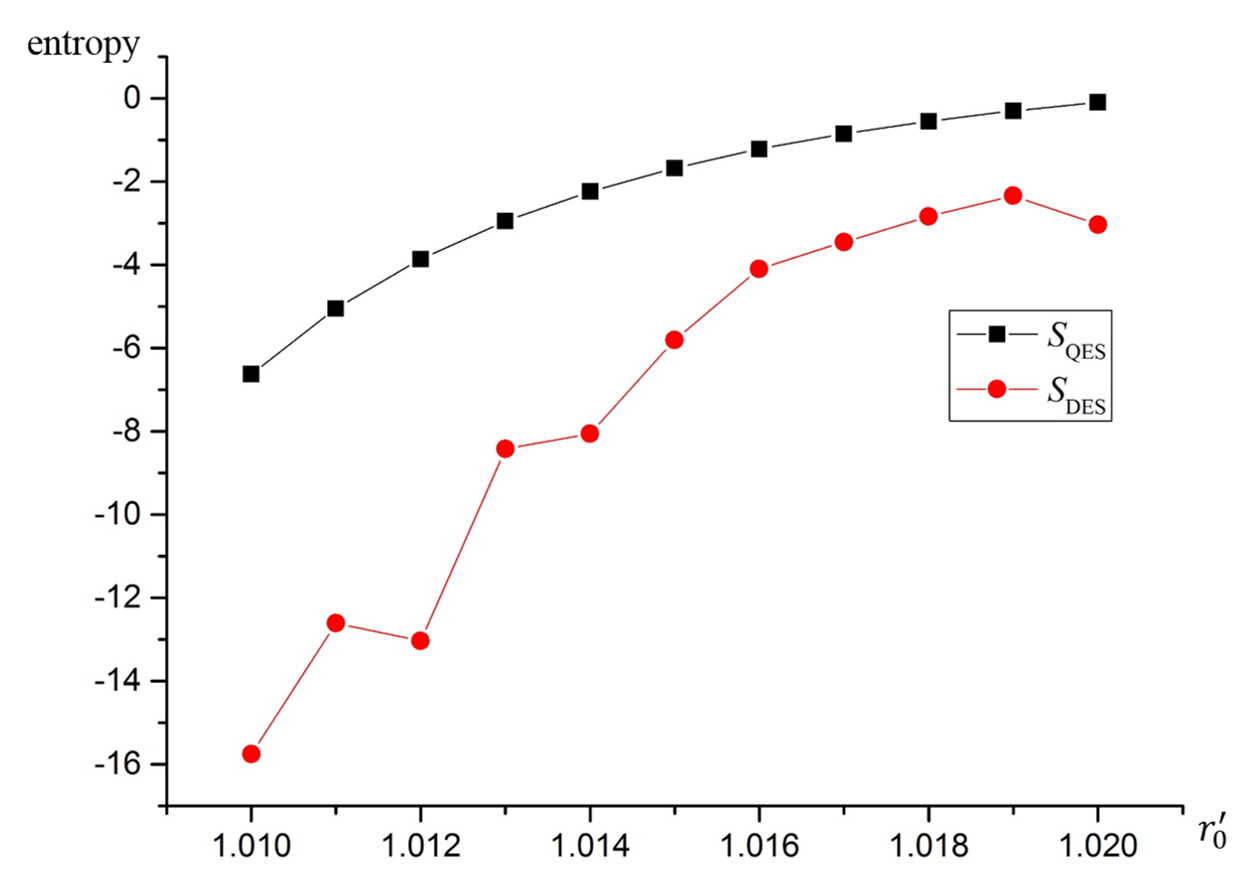}\\
  \caption{Plot of the entropy (subtracting the common UV divergent term (\ref{nispsph})) in the unit of $\frac{\pi l^3}{G_N^{(5)}}$ with respect to $r_0'$. We pick $\frac{1}{G_R^{(5)}}=\frac{2}{G_N^{(5)}}$ and $\theta=0.1$.}\label{DQES}
\end{figure}
\begin{figure}[h]
  \centering
  \includegraphics[width=9cm]{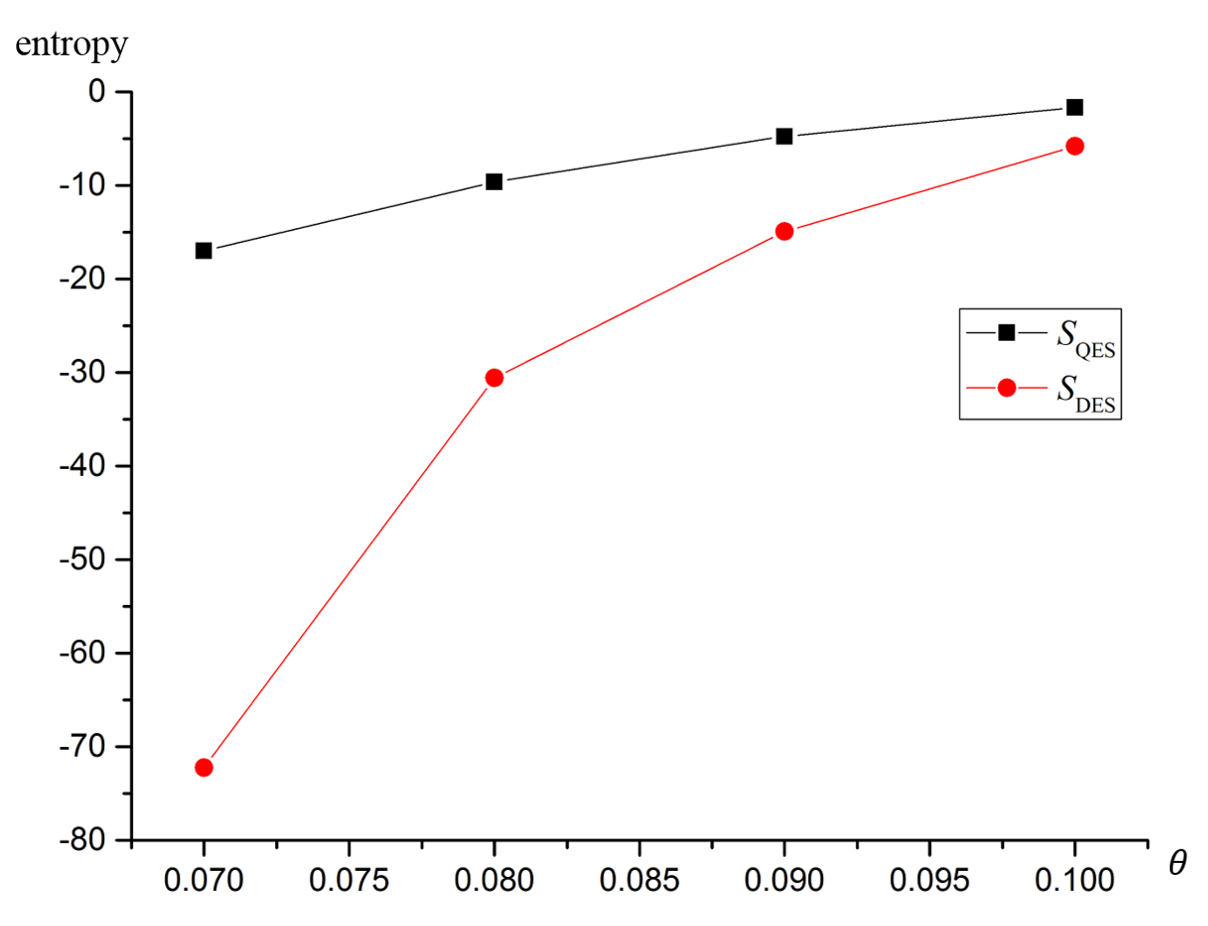}\\
  \caption{Plot of the entropy (subtracting the common UV divergent term (\ref{nispsph})) in the unit of $\frac{\pi l^3}{G_N^{(5)}}$ with respect to $\theta$. We pick $\frac{1}{G_R^{(5)}}=\frac{2}{G_N^{(5)}}$ and $r'_0=1.015$.}\label{DQEStheta}
\end{figure}
In this subsection we compare the entropy computed by DES formula and that by QES formula numerically. To proceed, we determine the cut-off $\epsilon_y$ in the DES side by demanding that the two formulae give the same extremal points (surfaces) on the brane. In other words, the extremal values of $r_1'$ are related to each other by (\ref{rdrq}), where $r'_Q$ is $r'_1$ in the QES subsection.

When defect extremal surface intersects with the brane, the generalized entropy (\ref{gensphDES}) gives the extremal condition
\begin{equation}
\begin{split}
\frac{dr'_1}{dr'_{1,Q}}S'_{\text{RT}}(r'_1)+\tilde{S}'_{\text{defect}}(r'_{1,Q})+\frac{\pi l^3}{2G_N^{(5)}}\left(-\frac{8r_{1,Q}'(1+r'^2_{1,Q})}{\epsilon_y^2\cos^2\theta(r'^2_{1,Q}-1)^3}-\frac{1+r'^2_{1,Q}}{r'_{1,Q}(1-r'^2_{1,Q})} \right)=0\ ,
\end{split}
\end{equation}
where the last term on the left side comes from the UV divergent term $S_{\text{defect}}^{(0)}(r'_{1,Q})$. Then by plugging in the extremal value of $r'_1$, one can solve that
\begin{equation}
\begin{split}
\epsilon_y=\left(\frac{\pi l^3}{2G_N^{(5)}}\frac{8r'_{1,Q}(1+r'^2_{1,Q})}{\cos^2\theta(r'^2_{1,Q}-1)^3}\frac{1}{\frac{dr'_1}{dr'_{1,Q}}S'_{\text{RT}}(r'_1)+\tilde{S}'_{\text{defect}}(r'_{1,Q})-\frac{\pi l^3}{2G_N^{(5)}}\frac{1+r'^2_{1,Q}}{r'_{1,Q}(1-r'^2_{1,Q})}}\right)^{\frac{1}{2}}\ .
\end{split}
\end{equation}
Inserting the solution of $\epsilon_y$ back in (\ref{DESsph}), we finally get the DES entropy. In Fig.\ref{DQES} and \ref{DQEStheta}, we plot the entropy achieved from the two formulae numerically. It can be seen that DES formula gives a smaller value.

\section{Conclusion and Discussion}\label{sec6}

In this paper we studied defect extremal surface in time dependent AdS/BCFT as well as in higher dimensions. Defect extremal surface as the holographic counterpart of the island formula in the context of static defect AdS/CFT, has been proposed in~\cite{Deng:2020ent}. In the present work we focus on the validity of defect extremal surface formula in dynamical cases and found that it gives the same Page curve as the boundary island formula in AdS$_3$/BCFT$_2$. The derivation relies on a decomposition procedure of the AdS bulk with a brane proposed in~\cite{Deng:2020ent}. An effective theory including both gravity region and flat space QFT naturally appears, because we do reduction for one part of the bulk using partial Randall-Sundrum and dualize the remaining part of the bulk by traditional AdS/CFT. In the present work, we extend the partial Randall-Sundrum+AdS/CFT procedure to higher dimensions and compare the entanglement entropy computed from bulk defect extremal surface and boundary island formula. Unlike the precise agreement found in $2d$, we found that defect extremal surface gives a smaller entropy in all cases we have checked in higher dimensions. We understand this as a consequence of the partial Randall-Sundrum we employ, which basically transforms the microscopic entropy such as some part of Ryu-Takayangagi surface, to the Bekenstein-Hawking area entropy and therefore increases it by some amount. In particular, these two entropies are the same in AdS$_3$/BCFT$_2$. We expect that a more refined reduction procedure, such as considering all Kaluza-Klein contributions in the partial reduction, will perhaps resolve the discrepancy. It is also interesting to consider the defect extremal surface as the UV completion of the island formula, which will be useful in clarifying the gravity/ensemble puzzle both in $2d$ and higher dimensions. We leave these questions for future work. 

\section*{Acknowledgements}
We are grateful for useful discussions with our group members in Fudan University. This work is supported by NSFC grant 11905033. YZ is also supported by NSFC 12047502,11947301 through Peng Huanwu Center for Fundamental Theory.

\appendix


\end{document}